\documentclass[useAMS,usenatbib,onecolumn]{mn2e}
\usepackage{subfigure}
\usepackage{graphicx}

\def\erf{\mathop{\rm erf}\nolimits}

\def\figscaleA{0.7}
\def\figscaleB{0.5}

\def\spose#1{\hbox to 0pt{#1\hss}}
\def\lta{\mathrel{\spose{\lower 3pt\hbox{$\mathchar"218$}}
     \raise 2.0pt\hbox{$\mathchar"13C$}}}
\def\gta{\mathrel{\spose{\lower 3pt\hbox{$\mathchar"218$}}
     \raise 2.0pt\hbox{$\mathchar"13E$}}}
\let\simless=\lta
\let\simgreat=\gta

\def\aj{AJ}
\def\apj{ApJ}

\def\mnras{MNRAS}

\title[Bar--Halo interaction]{The Bar--Halo Interaction--II. Secular
  evolution and the religion of N-body simulations}

\author[Weinberg \& Katz]{Martin D. Weinberg\thanks{E-mail:
    weinberg@astro.umass.edu (MDW); nsk@astro.umass.edu (NK)} and Neal
    Katz\footnotemark[1] \\ Department of Astronomy, University of
    Massachusetts, Amherst, 01003, USA}

\begin{document}

\label{firstpage}

\date{\today}
\pagerange{\pageref{firstpage}--\pageref{lastpage}} \pubyear{2005}

\maketitle
\begin{abstract}
  This paper explores resonance-driven secular evolution between a bar
  and dark-matter halo using N-body simulations.  We make direct
  comparisons to our analytic theory \citep{Weinberg.Katz:05a} to
  demonstrate the great difficulty that an N-body simulation has
  representing these dynamics for realistic astronomical interactions.
  In a dark-matter halo, the bar's angular momentum is coupled to the
  central density cusp (if present) by the Inner Lindblad Resonance.
  Owing to this angular momentum transfer and self-consistent
  re-equilibration, strong realistic bars {\it WILL} modify the cusp
  profile, lowering the central densities within about 30\% of the bar
  radius in a few bar orbits.  Past results to the contrary
  \citep{Sellwood:06,McMillan.Dehnen:05} may be the result of weak
  bars or numerical artifacts.  The magnitude depends on many factors
  and we illustrate the sensitivity of the response to the dark-matter
  profile, the bar shape and mass, and the galaxy's evolutionary
  history.  For example, if the bar length is comparable to the size
  of a central dark-matter core, the bar may exchange angular momentum
  without changing its pattern speed significantly.  We emphasise that
  this apparently simple example of secular evolution is remarkably
  subtle in detail and conclude that an N-body exploration of any
  astronomical scenario requires a deep investigation into the
  underlying dynamical mechanisms for that particular problem to set
  the necessary requirements for the simulation parameters and method
  (e.g. particle number and Poisson solver).  Simply put, N-body
  simulations do not divinely reveal truth and hence their results are
  not infallible.  They are unlikely to provide useful insight on
  their own, particularly for the study of even more complex secular
  processes such as the production of pseudo-bulges and disk heating.
\end{abstract}



\section{Introduction}
\label{sec:intro}

Current theoretical work in galaxy evolution attempts to relate the
epoch of galaxy formation to the state of galaxies today.  This
requires understanding an approximately ten gigayear interval over
which disk galaxies are in a slowly changing near equilibrium state.
Although this equilibrium will be punctuated by minor mergers, the
formation of bars, and the excitation of spiral structure, the disk's
existence tells us that these perturbations must be relatively mild.

As in Nature itself, cosmological simulations often produce barred
disks.  This has led to a cacophony of predictions about the
importance and implications of bars to the overall evolution of
galaxies in the presence of cuspy dark matter haloes.  To summarise,
theory predicts that bars can transfer angular momentum to haloes
through resonances and most groups agree that the bar does slow
\citep{Sellwood:81,Weinberg:85,Hernquist.Weinberg:92,Debattista.Sellwood:00,
Sellwood:03, Athanassoula:03,Holley-Bockelmann.Weinberg.ea:05}
although \citet[hereafter VK]{Valenzuela.Klypin:03} find a more modest
slow down. \citet{Debattista.Sellwood:00} use this and the
observational evidence that bars are not slow rotators to constrain
the density of the present day dark-matter halo.  \citet[hereafter
WK02]{Weinberg.Katz:02} argue that the angular momentum deposited in
the halo will change the halo density profile.  This latter work was
criticised by \citet{Sellwood:06} and \citet{McMillan.Dehnen:05} who
argue that the results of WK02 do not occur in there simulations.

The goal of this paper and the previous companion paper,
\citet[hereafter Paper I]{Weinberg.Katz:05a}, is an exposition of the
underlying dynamical principles that govern secular evolution and
their application to N-body simulations.  In many cases of interest,
the secular evolution is mediated by resonances and the implicit time
dependence of an evolving system affects the subsequent evolution.  We
use a numerical technique in Paper I for solving the perturbation
theory that allows arbitrary time dependence to be included.  In other
words, the history of the system does matter and time-asymptotic
results (e.g.  Landau damping or the Lynden-Bell \& Kalnajs 1972
theory, hereafter LBK\nocite{Lynden-Bell.Kalnajs:72}) do not give
accurate results \citep{Weinberg:04}.  Guided by the numerical
requirements derived in Paper I, we perform a series of simulations to
illustrate the special features of the bar-halo resonant interaction,
its dependence on the properties of the halo and the bar, and the
implication for N-body simulations to properly model this
interaction. We hope that our results will help to clarify and
reconcile some of the apparently disparate findings of other research
groups (op. cit.).  Most importantly, we show that the details of the
torque depend on all aspects of the interaction, the halo profile, the
bar shape, bar strength, bar pattern speed and history and cannot
simply be predicted by an application of the Chandrasekhar dynamical
friction or LBK formula.  For example, dynamical theory predicts that
a bar inside of a homogeneous core will not slow appreciably, which
contradicts the naive dynamical friction analogy; we also demonstrate
this using an N-body simulation below.

An N-body simulation adds two additional complications to the dynamics
of secular evolution.  First, an individual particle passing through a
resonance receives a perturbation that depends sensitively on its
initial position in phase-space.  The correct secular evolution is the
net average of many such trajectories.  In Nature, a dark matter halo
most likely has $N\rightarrow\infty$ from the perspective of any
simulation and, therefore, has no difficultly averaging over all of
phase-space.  The simulation, however, must have a sufficient number
of particles in the vicinity of the resonance to obtain the correct
net torque.  Paper I calls the resulting requirement on the number of
particles the \emph{coverage criterion}.  Second, representation of
the dark-matter and stellar components by an unnaturally small number
of particles leads to fluctuations in the gravitational potential.
For modern simulations, the magnitude of these fluctuations yields a
very long relaxation time but the interaction region for a resonance
has a very small phase-space volume.  Paper I shows that the noise is
sufficient to cause orbits to random walk through resonances.  Of
course, if some orbits walk out of the resonance, others walk in.
However for some resonances, ILR in particular, orbits should {\em
linger} near the resonance for many rotation periods. This increases
the amplitude and changes the dependence of the torque on the
phase-space distribution.  The fluctuation noise prevents this
lingering and in so doing changes both the amplitude of the net torque
and the location of the orbits in phase space receiving the torque.
Paper I shows that natural noise sources like satellites and subhaloes
will not destroy the lingering orbits.  The existence of multiple
regimes underlines the necessity of understanding the dynamical
mechanisms \emph{before} fully trusting the results of a simulation
constructed to investigate unknown dynamics. Paper I develops two
criteria, a {\em small-scale noise} particle number criterion that
treats the scale typical of gravitationally softened particles and a
{\em large-scale noise} criterion that describes scales typical of
basis expansions.

For simulations in this paper, we will use a basis expansion code
(also known by the Hernquist \& Ostriker 1992 moniker
\emph{self-consistent field (SCF) code}\nocite{Hernquist.Ostriker:92})
to solve the Poisson equation.  Our variant of this method is reviewed
in \S\ref{sec:method} along with the details of our initial conditions
and the bar perturbation.  We choose the expansion technique for three
reasons: 1) it is fast, scaling as ${\cal O}(N)$ with small overhead;
2) it restricts spatial sensitivity to the scales of interest; and 3)
it facilitates direct comparison to perturbation theory.  Paper I
shows that the coverage criterion and the small-scale noise criterion
dominates for the bar problem considered here.  The expansion
technique eliminates the small-scale noise and, therefore, in this
paper we are only able to address the breakdown of the coverage
criterion.  We encourage groups with particle-particle and
particle-mesh codes to test their codes as outlined in Paper I for the
effects of small-scale noise.  We investigate centring in
\S\ref{sec:dipole} and concur with \citet{Sellwood:06} and
\citet{McMillan.Dehnen:05} that a rotating quadrupole fixed to be
centred on one position leads to an $m=1$ artifact in the halo.  This
can be remedied by giving the bar a monopole component, i.e. a mass,
which allows it to establish its own centre naturally by conserving
linear momentum, removing the $m=1$ artifact.  In \S\ref{sec:fid}, we
simulate a bar's slow down, angular momentum transfer, and the
subsequent evolution of the dark matter halo using a large bar.  We
show that a strong Inner Lindblad Resonance (ILR) exists in a cuspy
dark-matter halo \citep[e.g.][hereafter NFW]{Navarro.Frenk.ea:97}.
The coupling at the ILR together with the self gravity of the affected
orbits drives the evolution of the inner cusp profile.  A large bar
decreases the particle number requirements derived in Paper I and,
because our inner dark-matter halo has a scale-free (power law)
profile, the same results should obtain with little dependence on the
bar size for an appropriately scaled bar mass.  We use this to
investigate and corroborate the predictions by decreasing and the size
and mass of the bar in \S\ref{sec:barsize}.  Then, in
\S\ref{sec:barshape} we investigate the dependence on bar shape and
pattern speed, and in \S\ref{sec:profiles} investigate the dependence
on the dark matter halo profile.  We compare with other published
findings in \S\ref{sec:others} and end with a discussion and summary
in \S\ref{sec:sum}.

\section{N-body method and models}
\label{sec:method}

\subsection{Potential solver}
\label{sec:expansion}

The disk and dark matter halo are evolved using a three-dimensional
self-consistent field algorithm \citep{Weinberg:99}.  This potential
solver uses an orthogonal function expansion to represent the density
and potential field.  Truncation of this expansion, then, limits the
spatial resolution scale.  These expansions are very efficient
computationally but are not adaptive.  In most N-body methods, either
the gravitational softening, introduced to decrease two-body
scattering, or the grid cell size determines the spatial resolution.
In such codes, it is the number of such spatial resolution elements
within the simulation volume that determines the effective number of
degrees of freedom, typically a very large number.  Expansion codes
limit the degrees of freedom, causing a large decrease in the
small-scale noise, i.e. two-body relaxation, making this class of code
ideal for simulating the long term evolution caused by resonant
dynamics
\citep{Earn.Sellwood:95,Clutton-Brock:72,Clutton-Brock:73,Kalnajs:76,Polyachenko.Shukhman:81,Fridman.Polyachenko:84,Hernquist.Ostriker:92,Hernquist.Sigurdsson.ea:95,Brown.Papaloizou:98,Earn:96}.
As demonstrated in Paper I, the relaxation times for this code, when
used with a radial and angular basis truncation sufficient for
resolving all the important resonances, are orders of magnitude longer
than a softened direct summation or tree code.  Conversely, if one
were to include enough radial and angular basis functions so that the
same spatial resolution as a softened particle code were approached,
the noise level and relaxation times would be comparable.  The choice
of a particular truncation introduces a bias by limiting the density
and potential profiles to those that can be represented by such a
basis.  Nonetheless, for near-equilibrium secular evolution, the
overall changes are likely to be on large spatial scales and to be
gradual and, therefore, less affected by incompleteness.

Our potential solver exploits properties of the Sturm-Liouville (SL)
equation to generate a numerical bi-orthogonal basis set whose lowest
order basis function matches the equilibrium model.  Many important
physical systems in quantum and classical dynamics reduce to the SL
form,
\begin{equation}
{{d}\over {dx}}{ \Big[ p(x) {{d\Phi(x)} \over {dx}} \Big] - 
  q(x) \Phi(x)} = {\lambda \omega(x)\Phi(x)}, 
\end{equation}
where $\lambda$ is a constant and $\omega(x)$ is a known function
called either the density or weighting function.  If $\Phi(x)$ and
$\omega(x)$ are positive in an interval $a<x<b$ then the SL equation
is satisfied only for a discrete set of eigenvalues $\lambda_j$ with
corresponding eigenfunctions $\phi_j(x)$ where $j=0,1,\ldots$.  The
eigenfunctions form a complete basis set \citep{Courant.Hilbert:53}
and can be chosen to be orthogonal with the following additional
properties: 1) the eigenvalues $\lambda_n$ are countably infinite and can
be ordered: $\lambda_n < \lambda_{n+1}$; 2) there is a smallest
non-negative eigenvalue, $\lambda_1 > 0$, but there is no greatest
eigenvalue; and 3) the eigenfunctions, $\phi_n$, possess nodes between
$a$ and $b$, and the number of nodes increases with increasing $n$,
e.g. the eigenfunction $\phi_1(x)$ has no nodes, $\phi_2(x)$ has one
node, etc.

In the special case of Poisson's equation, we use the eigenfunctions
to construct biorthogonal density and potential pairs, $d^{lm}_k$ and
$u^{lm}_j$ given by:
\begin{equation}
-{1\over4\pi G} \int dr r^2 d^{lm\,*}_k(r) u^{lm}_j(r) = {\delta_{jk}}.
\end{equation}
The lowest order potential-density pair ($j=k=1$, $l=m=0$) is defined
to be the equilibrium profile, and the higher order terms represent
deviations about this profile.  This approach minimises the number of
radial terms that one requires to reproduce both the unperturbed
equilibrium and large-scale variations from this equilibrium.
\citet{Weinberg:99} shows that by assuming a gravitational potential
of the form $\Phi(r)=\Phi_0(r)f(r)$ with physical boundary conditions
for Poisson's equation and $\Phi_0$ chosen to be the equilibrium
field, the equation for $f(r)$ also takes the SL form and can be
solved numerically to high accuracy using the Pruess \& Fulton algorithm
\citep{Marletta.Pryce:91,Pruess.Fulton:93,Pryce:93}.
\citet{Weinberg:99} suggested the possibility of reconstructing the
basis as the system evolves.  Although this procedure has some
advantages for studying long-term evolution, one must exercise great
care to ensure that transient features are not frozen into the basis.
For the simulations here, we fix the basis for the entire simulation.

We use a NFW profile truncated at large radii for our equilibrium halo model 
and as the zero-order radial basis function.  We obtain a self-consistent
phase-space distribution function for this model as follows.  We use
an Eddington inversion \citep[e.g.][]{Binney.Tremaine:87} to derive
the phase-space distribution function for the truncated profile,
$\rho_0(r)$ and $\phi_0(r)$, and integrate this distribution over
velocities to get $\rho_1(r)$ and $\phi_1(r)$.  We then repeat the
inversion to get a new distribution function and integrate to get
$\rho_2(r)$ and $\phi_2(r)$ and so on.  In practice, this procedure
converges in several iterations.  In many cases, we add an additional
spherical component to the gravitational potential that represents the
enclosed disk mass.  The initial conditions are generated from random
variates ${\cal R}$ by inverting the mass profile ${\cal R}=M(r)$ to
get a spatial position and by using the acceptance-rejection method
along with the phase-space distribution function to get a velocity.
Models constructed this way are usually in good virial equilibrium to
start, with $|2T/W+1|\simless0.01$ typically for $N=10^6$ particles.

\begin{figure}
  \centering
  \includegraphics[width=\figscaleB\linewidth]{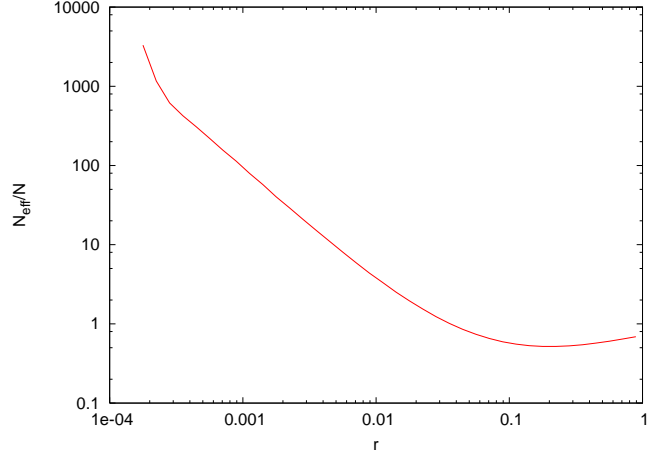}
  \caption{The ratio of the effective number of equal mass particles to 
  the actual number of particles as a function of halo radius for our multimass
  equilibrium. The
  ratio is near unity for $r\ga0.01$ but increases to almost 1000 for the
  most bound particles in the simulation.}
  \label{fig:multimass}
\end{figure}

To improve the effective particle number in the dynamically important
central density cusp, we generate initial conditions whose spatial
number density is $n(r)\propto r^{-2.5}$, steeper than the spatial
mass density of the inner NFW profile.  We do this by determining the
phase-space distribution function for this new number density profile
by Eddington inversion as described above.  The assigned mass is then
the mean mass per particle multiplied by the ratio of the {\em mass}
phase-space distribution function to the {\em number} phase-space
distribution function.  At any given radius, we can then define the
effective particle number, $N_{eff}(r) = M/\langle m\rangle_r$ where
$\langle m\rangle_r$ is the mean particle mass at radius $r$. The
ratio of $N_{eff}$ to the total particle number $N$ is shown in Figure
\ref{fig:multimass}.  The values of $N_{eff}$ in the cusp everywhere
exceeds $N$.  For example, an $N=10^6$ multimass particle phase space
is equivalent to a $N_{eff}=10^8$ equal mass particle phase space for
$r\le0.001$.

We retain halo basis terms up to $n_{\rm max}=20$ and $l_{\rm max}=4$.
Particles are advanced using a leapfrog integrator, with a time step
1/50 of the smallest orbital oscillation period.  Such a small time
step is required owing to the small interaction region around each
resonance.  This requirement varies with the bar model, as described
later.  We compared orbits integrated in the exact potential and using
the perturbation approach from Paper I with the results of the N-body
simulation to confirm our choice of time step and verify the
insignificance of noise.

\subsection{Details of the bar perturbation}
\label{sec:barpert}

We choose the bar to be a homogeneous ellipsoid with axes $a_1, a_2,
a_3$ and ratios $a_1:a_2:a_3::10:y:1$ with $2\le y\le7$.  As described
in \S\ref{sec:fid}, we choose our bar perturbation to be the monopole
and quadrupole parts of this potential.  The assumption of a
homogeneous ellipsoid does not limit the applicability of our work to
realistic bars owing to the weak dependence of the monopole and
quadrupole components to the details of the ellipsoid.  Rather, the
ellipsoid sets the overall scale.  The initial conditions are then
constructed by adding the monopole to the NFW profile and performing
the iterative Eddington inversion procedure.  This allows the NFW
density profile to be approximately maintained as the quadrupole part
of the perturbation is slowly turned on over several bar rotation
times.
  
The quadrupole is assumed to have the functional form:
\begin{equation}
  \Phi_1(r, \theta, \phi) = Y_{22}(\theta, \phi) U_{22}(r)
  \label{eq:quad}
\end{equation}
where
\begin{equation}
  U_{22}(r) = {b_1 r^2\over 1+(r/b_5)^5 }
  \label{eq:wkquad}
\end{equation}
This form has the correct solution to Laplace's equation at very small
and very large values of $r$ and is a good fit to the exact quadrupole
(see \S\ref{sec:HWWK}).  The derivation of $b_1$ and $b_5$ follows
from the exact potential in terms of elliptic integrals
\citep[see][]{Weinberg:85}.  $U_{22}$ peaks at $r=(2/3)^{1/5}b_5$ with
the value $(2^23^3)^{1/5}b_1 b_5^2/5$.  For this potential, the
projected surface density along the minor axis is
\begin{equation}
        \Sigma(x,y) = {3M_b\over 2 \pi a b} \sqrt{1 - x^2/a^2 - y^2/b^2}
        \label{eq:barsurf}
\end{equation}
where $M_b$ is the total bar mass.  This form approximately describes
the observed surface brightness of bars \citep{Kormendy:82}.  However,
alternative profiles can be fit with equation (\ref{eq:wkquad}) or the
more general form below (eq. \ref{eq:genquad}).

\subsection{Comparison with bar quadrupole from Hernquist \& Weinberg}
\label{sec:HWWK}

\begin{figure}
  \centering
  \includegraphics[width=\figscaleB\linewidth]{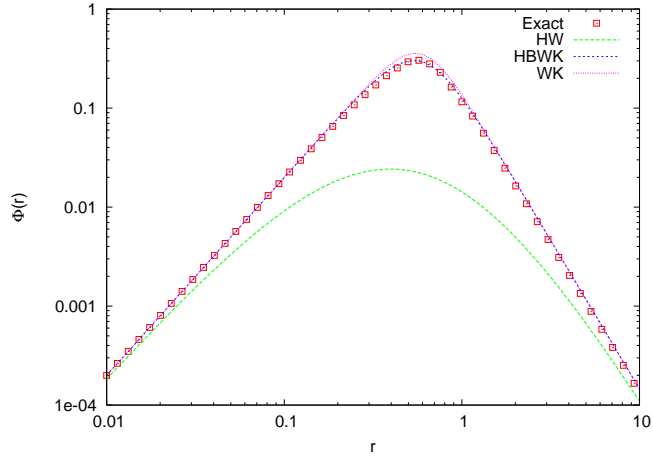}
  \caption{Comparison of the quadrupole potential fitting formula from
    HBWK, WK, and HW with the exact quadrupole potential of the
    homogeneous ellipsoid whose projected surface density is given in
    eq. \protect{\ref{eq:barsurf}}. Notice that the HBWK bar, from a
    fully consistent disk + halo N-body simulation is
    indistinguishable from the exact homogeneous ellipsoid at
    quadrupole order. }
  \label{fig:potcof}
\end{figure}

\begin{figure}
  \centering
  \includegraphics[width=\figscaleB\linewidth]{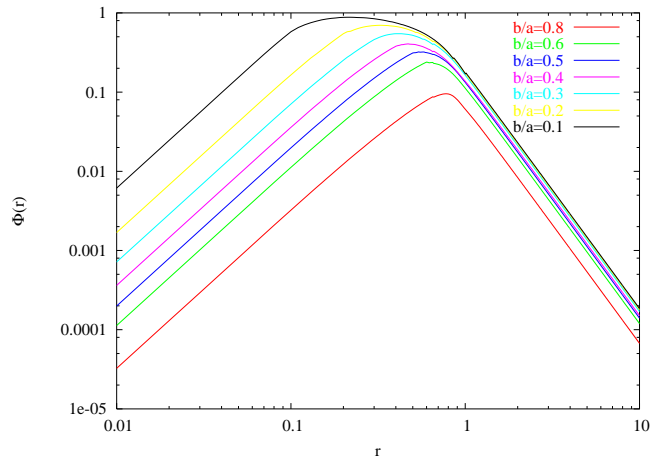}
  \caption{Comparison of the WK quadrupole potential fitting formula
    for bars with various axis ratios $b/a$ with fixed
    $c/a=1/20$.  A smaller $b/a$ yields a deeper quadrupole potential
    inside the bar radius,  which increases the torque for small
    energy cusp orbits.}
  \label{fig:quadcomp}
\end{figure}

We used the quadrupole form of equation (\ref{eq:wkquad}) in WK, but
a different form,
\begin{equation}
  U_{22}(r) = {b_1 r^2\over \left[ 1+r/b_5\right]^5},
  \label{eq:hwquad}
\end{equation}
in \citet[hereafter HW]{Hernquist.Weinberg:92}.  Both of these can be
included in the more general form
\begin{equation}
  U_{22}(r) = {b_1 r^2\over \left[ 1+(r/b_5)^\alpha\right]^{5/\alpha}}
  \label{eq:genquad}
\end{equation}
while still maintaining the same asymptotic radial dependence
consistent with the Laplace equation.  Equation (\ref{eq:hwquad})
results for $\alpha=1$ and equation (\ref{eq:wkquad}) results for
$\alpha=5$.  Figure \ref{fig:potcof} compares these two fitting
formula to the exact potential of a homogeneous ellipsoid.  In both
cases $b_1$ and $b_5$ are chosen to match the exact potential at small
and large $r$.  \citet[hereafter
HBWK]{Holley-Bockelmann.Weinberg.ea:05} find stellar bars, formed
self-consistently in N-body simulations, have $\alpha\approx4$.  As
shown in the figure, the quadrupole from the ellipsoid is a very good
match to the quadrupole derived from the N-body bar.  Equation
(\ref{eq:wkquad}) nearly matches the exact potential for the ellipsoid
while equation (\ref{eq:hwquad}) underestimates the peak of the
potential by more than an order of magnitude.  This difference would
have a significant effect on the total torque.

Different bar profiles may require tuning the value of $\alpha$.
Increasing values of $\alpha$ in equation (\ref{eq:genquad}) increases
the sharpness of the turn over between the $r^2$ and $r^{-3}$ tails; the
profile becomes a discontinuous broken power law as
$\alpha\rightarrow\infty$.  \citet{Kormendy:82} reports that along the
major axis the surface brightness is nearly constant interior to a
sharp outer edge, while along the minor axis the profile is steep, as
in an $r^{1/4}$ law.  The approximation $\Sigma(x, y)$ above is,
therefore, a good representation of the major axis although it falls
off more gradually along the projected minor axis.  One can compensate for this
by decreasing the ratio of $a_2/a_1$.  Moreover, our fiducial choice of
$a_2/a_1=1/5$ (see \S\ref{sec:barpert}) is similar to those of many
bars \citep{Kormendy:82}.  Figure \ref{fig:quadcomp} shows that a
smaller $a_2/a_1$ gives a larger amplitude quadrupole inside the bar
radius, which will increase the efficiency of coupling to lower energy
orbits.

\section{Dipole response and the expansion centre}
\label{sec:dipole}

Application of equation (\ref{eq:quad}) as a bar perturbation requires
a prior knowledge of the galaxy's centre.  Although our simulations
contain sufficient particles such that the root mean square drift of the
centre will be negligible, inclusion of the $l=1$ term in the
potential solver with a fixed-centre quadrupole promotes an
instability between the bar and halo, which changes the angular momentum
exchange by amplifying the offset of their two centres.  A small shift
between the quadrupole and halo centre, which could be generated by particle
noise, converts some of the bar's angular momentum into linear momentum
of the halo cusp.  This increases the $l=1$ amplitude and leads to a
run away.  The consequences of this were noted by \citet{Sellwood:03}.
This instability may be removed by adding the monopole term to
equation (\ref{eq:quad}):
\begin{equation}
  \Phi_1({\bar r}, {\bar\theta}, {\bar\phi}) = 
  Y_{00}({\bar\theta}, {\bar\phi}) U_{00}({\bar r}) + 
  Y_{22}({\bar\theta}, {\bar\phi}) U_{22}({\bar r})
  \label{eq:monoquad}
\end{equation}
where ${\bar{\bf r}}$ is the coordinate the bar frame and $U_{00}$ is
the monopole part of the bar potential.  We compute the origin of the
bar frame by conserving linear momentum in the combined bar--halo
mass distribution.  The restoring force resulting from the addition of
inertial and gravitational mass causes the bar to oscillate weakly
about the centre and damps the $m=1$ instability.  We have performed
simulations both without the $l=1$ term and using only the even $l$ terms and,
when one includes the monopole part of the bar potential,
we obtained nearly identical torque curves $L_z=L_z(t)$ as when all the terms
are included.
In most cases, we choose the monopole and quadrupole to be consistent with a
single homogeneous ellipsoid, however, this is not a requirement.  As
long as the $m=1$ instability does not grow, the quadrupole amplitude
can be adjusted independently of the monopole to mimic a bar that
lengthens and changes shape.  

By adding the monopole component to the bar potential we successfully damped
the $m=1$ instability. However, since the bar centre and central density
cusp can now be offset, the central density cusp is no longer necessarily centred
on the Poisson expansion, which could introduce another problem.  If the offset
from the expansion centre became large enough then our finite number of 
expansion terms would cause the gravitational forces within the cusp to be 
underestimated and the cusp could artificially dissolve.
We check the position of the cusp relative to the expansion origin by 
computing the centre of mass of the most bound particles such that
we include a mass fraction of $5\times10^{-4} M_{vir}$.
For example, for $10^6$ equal mass particles within the virial radius,
we compute the centre for the 500 most bound particles.
Even if the bar is free to move relative to the density cusp, the typical
offset of the density cusp from the expansion centre
is only several hundredths of a percent of the virial radius. 
For the number of terms we keep in our Poisson expansion, i.e. $n_{max}=20$
and $l_{max}=4$, the central cusps of equilibrium NFW halos evolved with
offsets from the expansion centre of this size remain in equilibrium for over 
5 Gyrs, mitigating any need for concern.

We choose the pattern speed of the bar, $\Omega_p$, to be the circular
frequency in the unperturbed halo at some multiple of the bar length,
$a_1$.  The angular momentum of the bar is then:
\begin{equation}
  L_{z,bar} =  \Omega I_z,
\end{equation}
\begin{equation}
  I_z = {1\over5}(a_1^2 + a_2^2) M_b.
\end{equation}
The pattern speed at any future time is determined directly from
momentum conservation for the combined bar--halo system as follows.
Initially, we compute the total angular momentum content of the halo
($L_{z,halo}$), add this to the bar angular momentum, ($L_{z,bar}$),
and keep this constant.  The pattern speed is then
\begin{equation}
  \Omega_p = (L_{z,bar} + L_{z,halo} - L_{z,t})/I_z
  \label{eq:momcon}
\end{equation}
where $L_{z,t}$ is the angular momentum of the halo at time $t$.  This
model, of course, neglects angular momentum transport between the bar
and the rest of the disk; this is explicitly investigated and
quantified in HBWK.

\section{Fiducial run}
\label{sec:fid}

Our fiducial run uses an NFW halo with a concentration parameter of
$c=15$.  Our fiducial bar has a length equal to the NFW scale length $r_s$, 
a mass equal to 1/2 the mass of the halo enclosed within that radius, and
a shape $a_1:a_2:a_3::10:2:1$.
The perturbation is turned on over approximately a bar rotation time.
The bar pattern speed is adjusted during the simulation to conserve
the total angular momentum of the bar--halo system
(eq. \ref{eq:momcon}).  This bar radius is unrealistically large
compared to present-day bars, but this choice
ensures that all of the particle number criteria obtained in Paper I
are satisfied.  However, because the NFW profile is scale free inside
of $r_s$, one should not expect significant differences for smaller
bars scaled to the same enclosed dark-matter mass.  We will discuss
smaller present-day-sized bars in
\S\ref{sec:barsize}.

The bar perturbation is turned on over several bar rotation times
using an error function as follows:
\begin{equation}
A(t) = {A_o\over 2}\left\{1 + \erf[(t-t_0)/\delta]\right\}
\label{eq:turnon}
\end{equation}
where $A(\infty)=A_o$ is the final quadrupole amplitude.  The
amplitude of the monopole term, now integral to the overall
equilibrium, is not adjusted.  For the simulations described below, we
set $t_0=1/2$ and $\delta=1/4$ unless otherwise stated.  In Paper I we
used a slower turn on to investigate the detailed dynamical mechanism.
Here, our turn on rate is motivated by bar formation times observed in
self-consistent simulations (HBWK).  Scaled to the Milky Way, one time
unit is approximately 2 Gyr. Variations in $t_0$ and $\delta$ or the
choice $A(t)=\hbox{constant}$ do not make large differences in the
results since the strongest coupling is driven by the slowing of the
bar \citep{Weinberg:04}. To compare models with different parameters
$t_0$ and $\delta$, we define a scaled time $\tau\equiv\int^t_0
dt\,A(t)$.  This is the equivalent time for a full-strength bar.  In
an N-body simulation without the bar perturbation, the dark matter
halo profile is indistinguishable from the initial profile after 3
time units (approx.  6 Gyr for the Milky Way).

We use $10^6$ multimass particles for investigating the large bars
with fiducial parameters but smaller bars require larger $N$ as we
will see in later sections.  The criteria calculations from Paper I
predict that $N\approx10^8$ equal mass particles within the virial
radius are needed for ILR and $N\simless10^6$ equal mass particles for
the other low-order resonances.  The effective particle number in the
vicinity of ILR (see Paper I) for $N=10^6$ is between $10^7$ and
$10^8$ equal mass particles within the virial radius and this provides
sufficient coverage for all but possibly the most eccentric orbits
near the ILR for our large fiducial bar.  However, this particle
number is not sufficient for a scale-length sized bar, which requires
$N\simgreat10^9$ equal mass particles.

\subsection{Description and theoretical interpretation}

\begin{figure}
\centering
  \includegraphics[width=\textwidth]{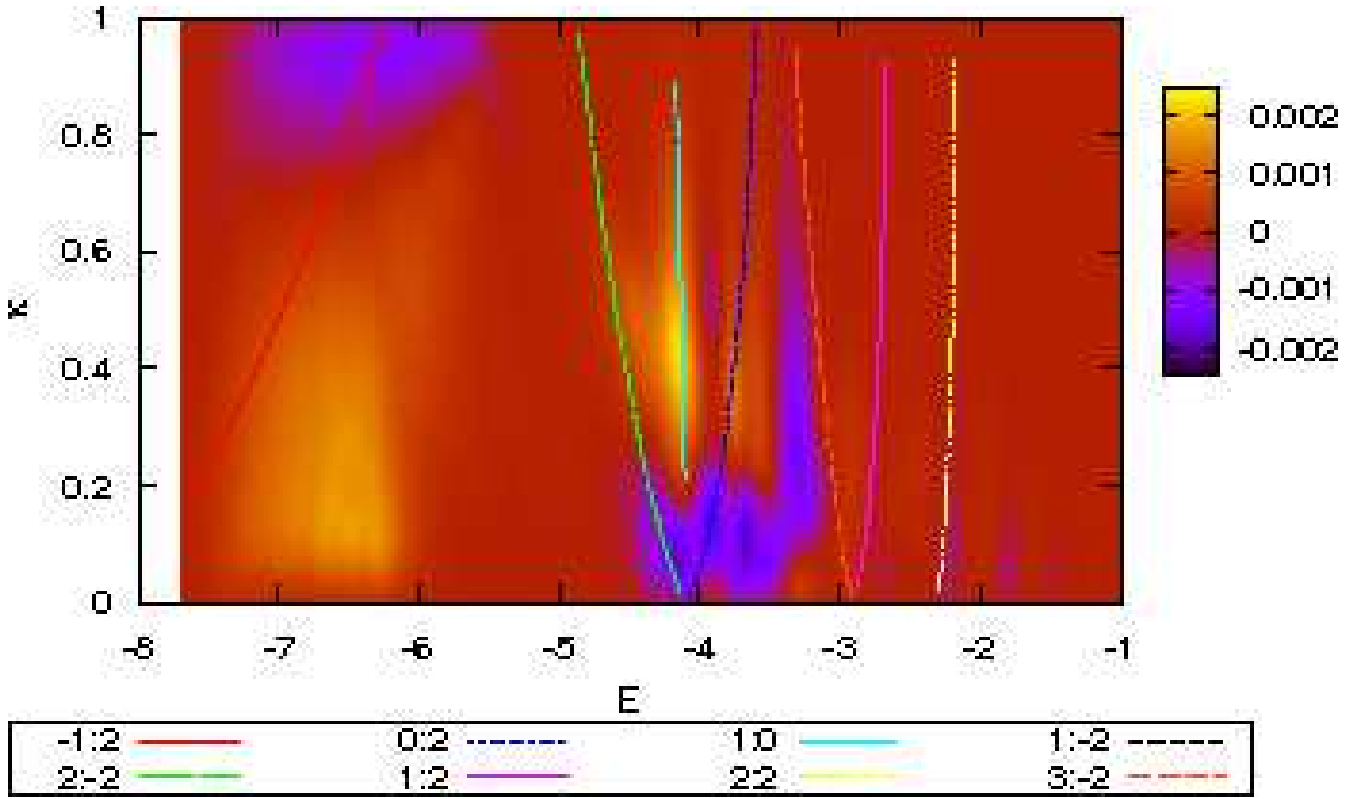}
\caption{Distribution of $\Delta L_z$ for the slowing fiducial bar
  with the locations of low-order resonances $l_1:l_2$ at the pattern
  speed of peak torque overlayed.}
\label{fig:dlz}
\end{figure}

The resonant angular exchange mechanism deposits (or extracts) angular
momentum in (from) the dark matter halo at specific regions in phase
space.  Hence, as we discuss in Paper I, plotting the change in
angular momentum over a finite time is a good way to see the important
resonances.  In Figure \ref{fig:dlz}, we show the ensemble change in
the $z$ component of the angular momentum $\Delta L_z$ in phase-space
during the evolution of the fiducial bar.  An equilibrium phase-space
distribution in a spherical halo is described by two conserved
quantities and for a clearer presentation we use the energy $E$ and
the angular momentum scaled to the maximum for a given energy
$\kappa\equiv J/J_{max}(E) \in [0,1]$.  This figure is made by first
computing $\Delta L_z$ for each orbit as a function of its initial
values of $E$ and $\kappa$.  We then use kernel density estimation
with cross validation \citep{Silverman:86} to estimate the smoothing
kernel.  We increase and decrease this estimate to ensure that the
resulting density field is not over smoothed. In the figure, we also
indicate the positions of the low-order resonances, calculated at the
bar pattern speed when the torque is largest. However, since the bar
pattern speed changes with time so will the position of the resonances
in phase space. This makes the ILR less obvious because it is spread
over a large range in energy owing to the time-dependence of the
frequency spectrum of the slowing bar.  This resonance, $(-1,2, 2)$
\footnote{The resonance condition may be written
$l_1\Omega_r+\l_2\Omega_\phi = m\Omega_p$ for a spherical model.  We
denote a particular resonance by the triple $(l_1, l_2, m)$. When we
are considering $m=2$ specifically, we shorten the designation to
$(l_1, l_2)$.}, has a particularly interesting degeneracy. As
$\Omega_p\rightarrow0$, there is always some value
$\kappa\rightarrow0$ such that $l_1\Omega_r +
l_2\Omega_\phi\rightarrow m\Omega_p$.  The limiting case is a purely
radial orbit.  In a cuspy profile, the orbital frequencies $\Omega_j$
increase as the energy decreases so that there is also some bit of
phase space near resonance.  This resonance track continues to smaller
values of $E$ and $\kappa$.  Unlike the non-degenerate low-order
resonances, for ILR a small change in $\Omega_p$ can have a large
effect on the relative location of the resonance track.

\begin{figure}
  \includegraphics[width=\figscaleA\linewidth]{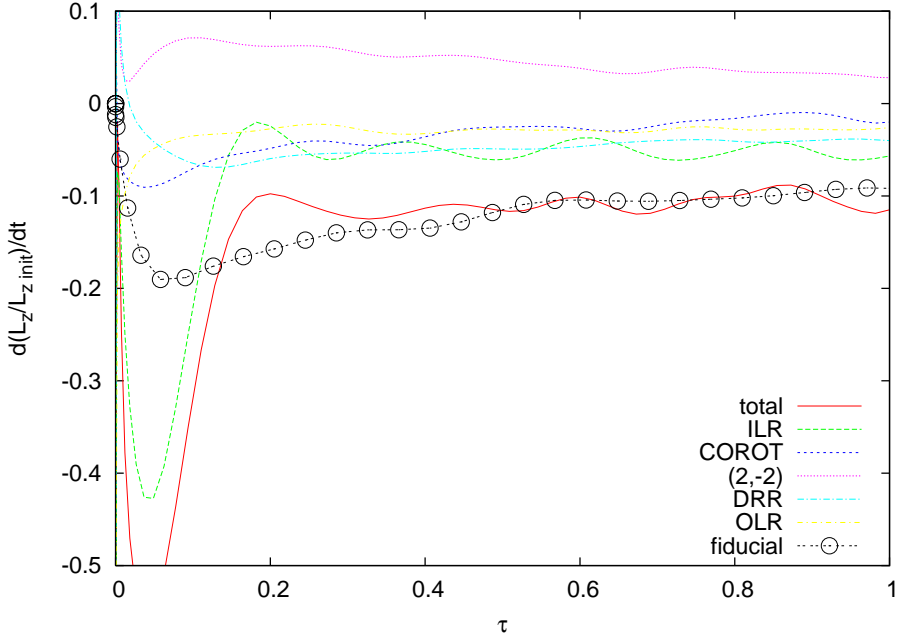}
  \caption{The torque $\Delta L_z/\Delta t$ versus scaled time
    $\tau$ from the fiducial model compared with the perturbation
    theory predictions from Paper I for the total torque and for the
    torque from individual resonances.}
  \label{fig:dlz_compare}
\end{figure}

In an evolving system, the frequency spectrum is broadened, and this
promotes coupling to the ILR over a large range of radii.  Paper I
describes the application of canonical perturbation theory to
time-dependent secular evolution using a numerical procedure.  The
perturbation theory allows investigation of each resonance
separately. We compare the perturbation theory predictions to our
fiducial bar simulation in Figure \ref{fig:dlz_compare}.  The
perturbation theory predicts that the ILR is responsible for
approximately half of the total torque during the rapid phase of bar
slowing.  However, since the strength of the bar perturbation causes
the interaction to be nonlinear, the linear perturbation theory is
scaled beyond its domain of validity, which results in the mismatch at
early times ($\tau\simless0.2$).  Because we do not expect the
perturbation theory to match the stronger bar simulation precisely as
it did for the weak bars considered in Paper I, the exact comparison
is less useful here than are the relative contributions of each
resonance.  After the initial nonlinear phase, the ILR, the direct
radial resonance $(1,0,2)$ [hereafter, DRR], and the $(2,-2, 2)$
resonances account for approximately 25\% of the total torque each and
corotation and OLR are 15\% and 10\% of the total, respectively.  Also
note that the $(2,-2,2)$ resonance \emph{supplies} angular momentum to
the bar.  After the bar slows substantially, the ILR becomes a less
efficient angular momentum sink, requiring increasingly
high-eccentricity orbits for coupling to the cusp and the remaining
low-order resonances move to large radii where the coupling is also
weak.  As we show in Paper I, the ILR is the most numerically
difficult resonance to reproduce in N-body simulations.  If numerical
deficiencies cause the ILR to be missed or weakened, the net torque on
the bar can be greatly reduced.  Since, of the remaining resonances,
DRR and $(2,-2,2)$ have comparable strengths but opposite signs and
effect similar radii, in this numerically compromised case only the
net effect of corotation and OLR will slow the bar and the angular
momentum will be deposited at larger radii in the halo, at about the
bar radius.

\begin{figure}
  \includegraphics[width=\figscaleA\linewidth]{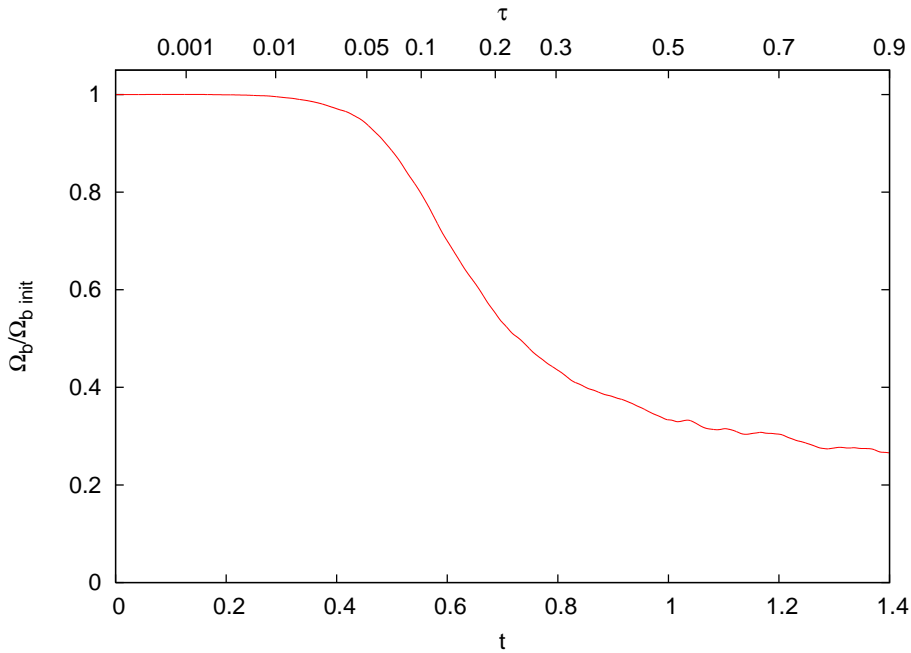}
  \caption{Evolution of the fiducial bar's pattern speed in physical
  time $t$ (lower axis) and scaled time $\tau$ (upper axis).}
  \label{fig:fidevol}
\end{figure}

\begin{figure}
  \subfigure[Evolving pattern speed]{
    \includegraphics[width=0.5\linewidth]{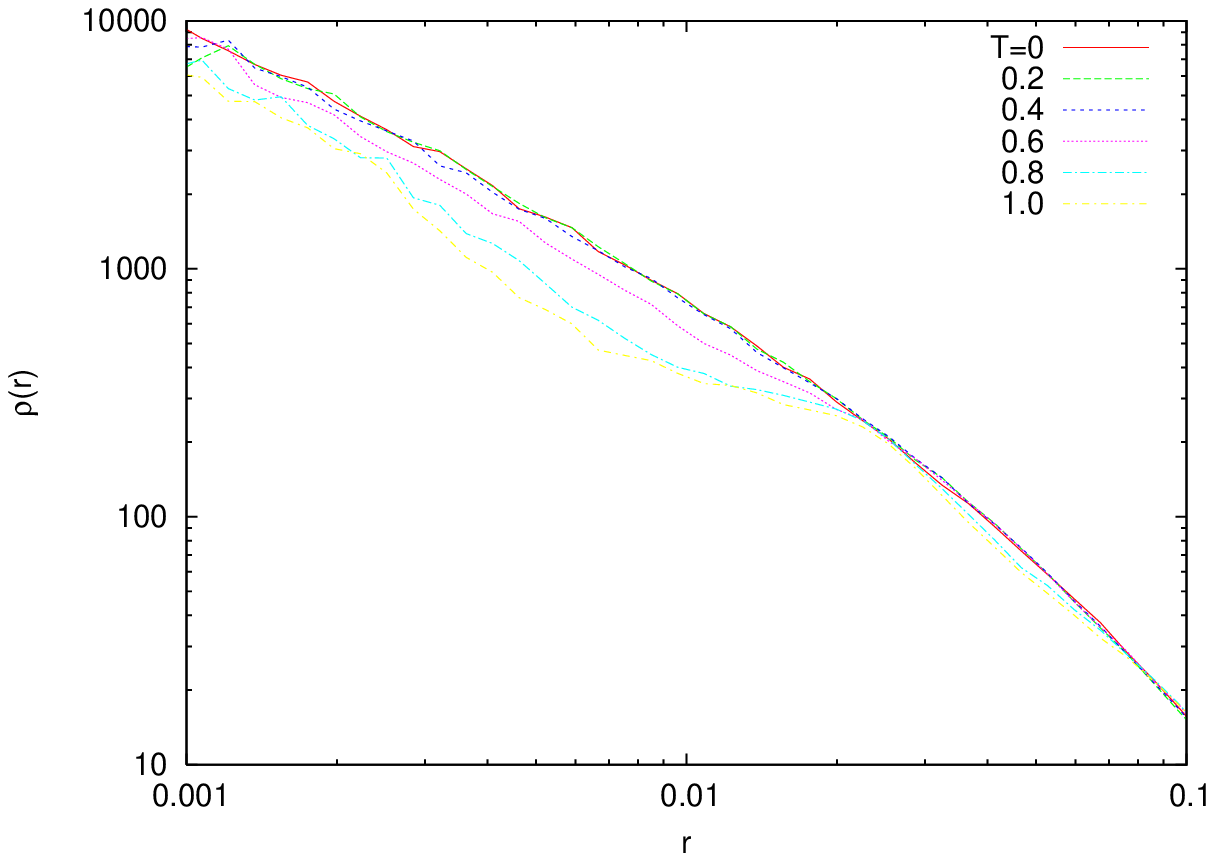}
  }
  \subfigure[Fixed pattern speed]{
    \includegraphics[width=0.5\linewidth]{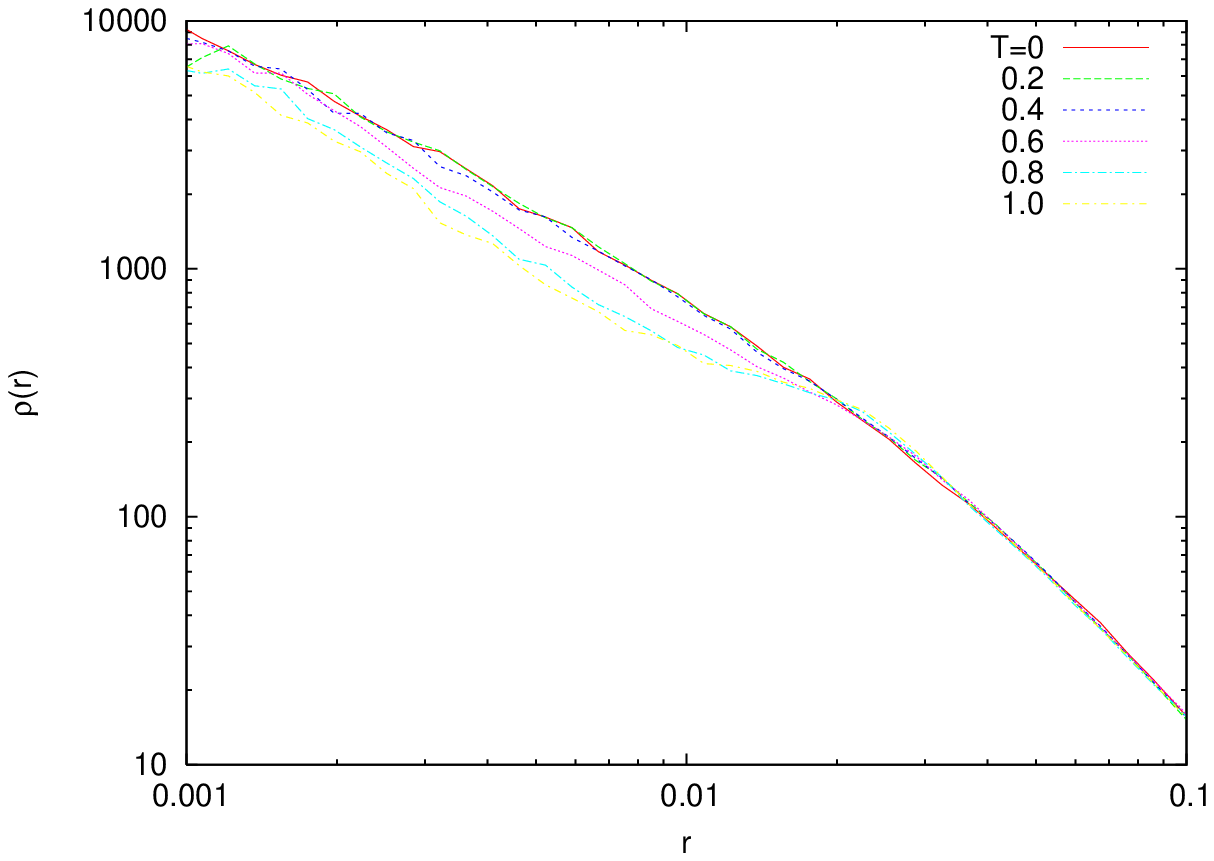}
  }
  \caption{Evolution of the halo density profile shown as function of time $T$.
    The initial bar period is 0.3 time units.}
  \label{fig:profiles}
\end{figure}

We plot the bar pattern speed evolution in Figure
\ref{fig:fidevol}. The initial bar period is $0.3$ time units and the
bar reaches half of its full amplitude at $t=0.5$ ($\tau=0.07$).  At
about this time the bar begins to slow rapidly.  As described above,
the angular momentum loss is dominated by the ILR $(-1,2, 2)$ and DRR.
As we discussed in WK02, the addition of this angular momentum causes
the halo density profile to evolve.  We show the bar-driven evolution
of the dark matter density profile for the fiducial model in Figure
\ref{fig:profiles}a.  The orbits affected by the ILR have
characteristic radii well inside the bar radius.  The corotation
radius begins at the end of the bar and slowly moves outward as the
bar slows, while the ILR moves inwards.  The DRR occurs close to but
inside of corotation.  The initial bar radius is $r_s = 0.067$ and the
peak of the profile evolution occurs at a radius of 0.01, a factor of
six smaller in radius.  Note that this profile evolution occurs even
though we added a monopole component to the bar potential and hence
removed the $m=1$ instability.

Even if the bar does not slow at all the halo density profile still
evolves.  We plot the evolution of the profile for this
fixed-pattern-speed case in Figure \ref{fig:profiles}b and it shows
the same overall trend as for the evolving bar.  However, the
evolution in the fixed-pattern case is driven only by the evolving
halo and the creation of the bar and not by the pattern speed
evolution and, therefore, is weaker.  For times longer than $t\ga1$,
the evolution of the inner profile ceases for our fiducial bar, since
the bar has mostly stopped slowing at this time.  We restarted a fresh
bar in the evolved phase space at $t=1$ and the evolution continued
for another time unit. This indicates that the finite life time of the
bar is spreading power to lower frequencies, where it can be
resonantly absorbed by the halo cusp; we have predicted this transient
analytically using the methods of Paper I.

\subsection{Profile evolution explained}

\begin{figure}
  \begin{minipage}{0.5\textwidth}
    \includegraphics[width=\linewidth]{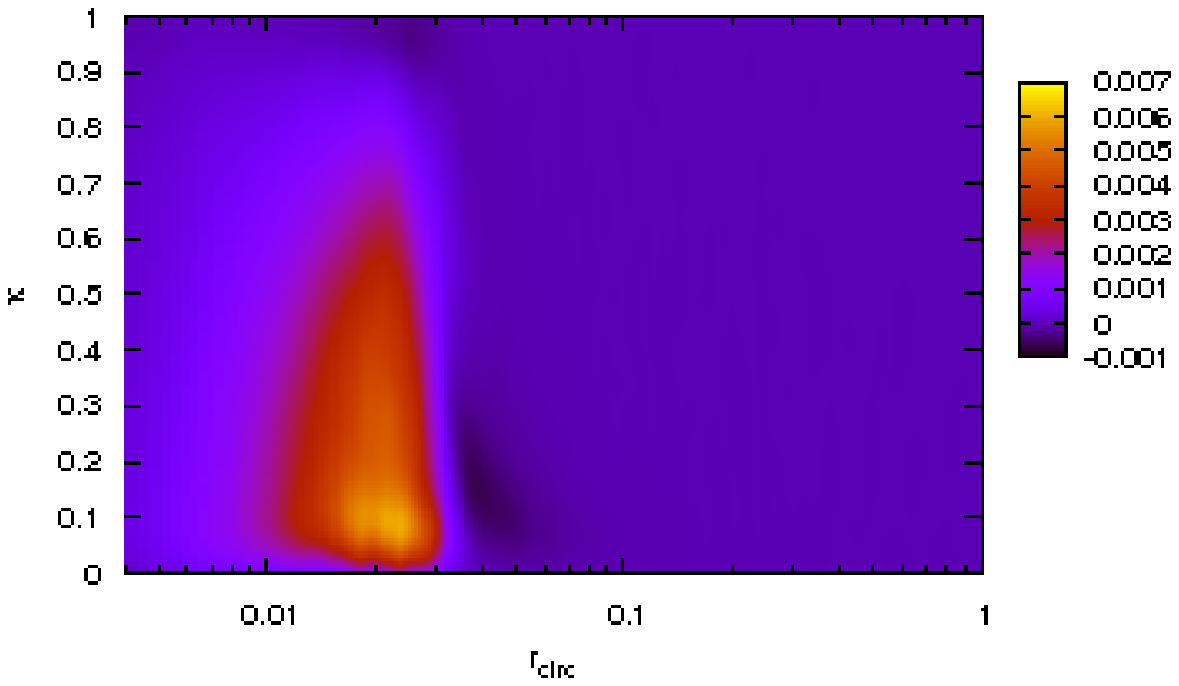}
    \caption{The distribution of z-angular momentum change 
      ($\Delta L_z$) in the halo
      phase space plotted as a function of $r_{circ}$ and
      $\kappa=J/J_{max}$ predicted by perturbation theory.
      The quantity $r_{circ}$ is the radius of a circular
      orbit with the same energy.}
    \label{fig:fid_ilr_dlz}
  \end{minipage}
  \hfill
  \begin{minipage}{0.45\textwidth}
    \includegraphics[width=\linewidth]{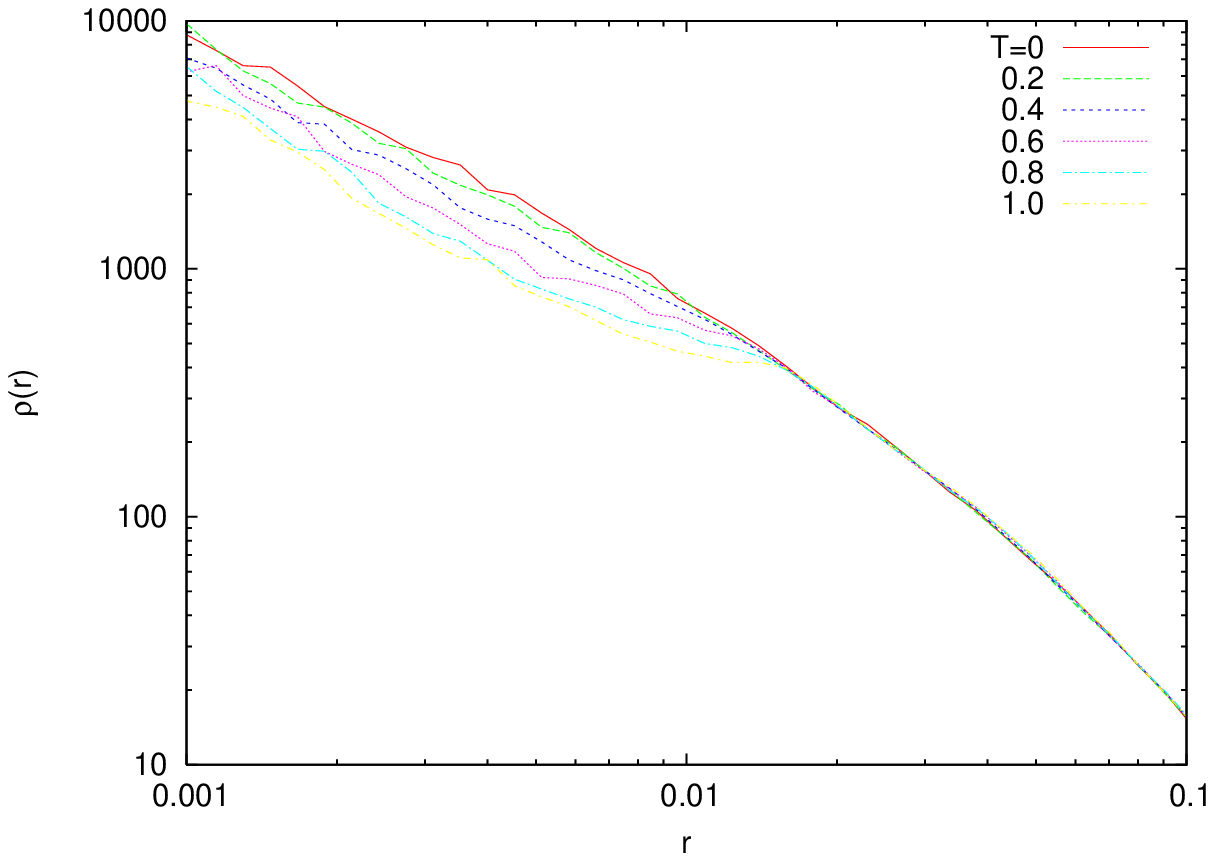}
    \caption{The evolution of the halo profile shown as a function of time $T$
      after adding the distribution of $\Delta L_z$ shown in Figure
      \protect{\ref{fig:fid_ilr_dlz}} over one time unit.}
    \label{fig:dlz_profiles}
  \end{minipage}
\end{figure}

Figure \ref{fig:fid_ilr_dlz} shows the phase-space distribution of
angular momentum change $\Delta L_z$ during one time unit (in units
with $M_{vir}=R_{vir}=1$) for the ILR as the bar slows.  This
distribution was computed using the perturbation theory approach used in
Paper I and plotted as in Figure \ref{fig:dlz}.
Rather than plotting energy $E$ and scaled angular momentum $\kappa\equiv
J/J_{max}$, however, we plot the radius of the circular orbit
with the same energy $E$, called $r_{circ}$.  For $\kappa\sim1$,
the radius of the actual orbit will be well approximated by $r_{circ}$,
but for $\kappa\sim0$, the pericentric radius will be much smaller
than $r_{circ}$, making the apocentric radius ${\cal
O}(r_{circ})$.  Since the largest values of $\Delta L_z$ occur at
small values of $\kappa$, the torqued radial orbits have an effect
on the cusp well inside $r_{circ}$.  Of course, to follow the effect on
the new equilibrium profile one must allow the phase space
to self consistently readjust to the new equilibrium.
Comparing Figures \ref{fig:profiles}
and \ref{fig:fid_ilr_dlz} shows that the radii that have the greatest profile
evolution are the same radii with the greatest angular momentum change,
suggesting that the profile evolution owes to angular momentum transfer 
at ILR.

As further evidence, we perform a simulation in which we add the 
perturbation theory predicted change in angular
momentum (from Figure \ref{fig:fid_ilr_dlz}) at each point in phase space
to an equilibrium NFW halo, continuously over one time unit and allow
the system to reach a new equilibrium. More precisely,
at each time step we use Figure \ref{fig:fid_ilr_dlz} to
estimate the time-averaged torque $T_z=\Delta L_z/\Delta t$, 
then we compute the values of $E$ and $\kappa$ for each
particle, then we compute $T_z$ by interpolation, and finally we accelerate the
particle in azimuth to reproduce $\Delta L_z=T_z\Delta
t$.  The resulting density profiles are shown in Figure
\ref{fig:dlz_profiles} and are a very close
match to those in Figure \ref{fig:profiles}a, demonstrating
that the deposition of angular momentum at ILR and subsequent
re-equilibration drives the profile evolution. In principle, any
torque received by the halo must change the underlying equilibrium
profile.  The change in the cusp appears dramatic because of the low overall 
specific angular momentum of the ILR-coupled cusp orbits relative to
the bar angular momentum and the crucial role these orbits play in sustaining
the self gravity of the cusp.  The response would presumably be
smaller for a steeper cusp, which would have a larger fractional binding
energy.

\section{Dependence on model parameters}
\label{sec:ncrit}

\subsection{Variation of bar size and bar mass}
\label{sec:barsize}

\begin{figure}
  \subfigure[Fast-limit scaling]{
    \includegraphics[width=0.49\linewidth]{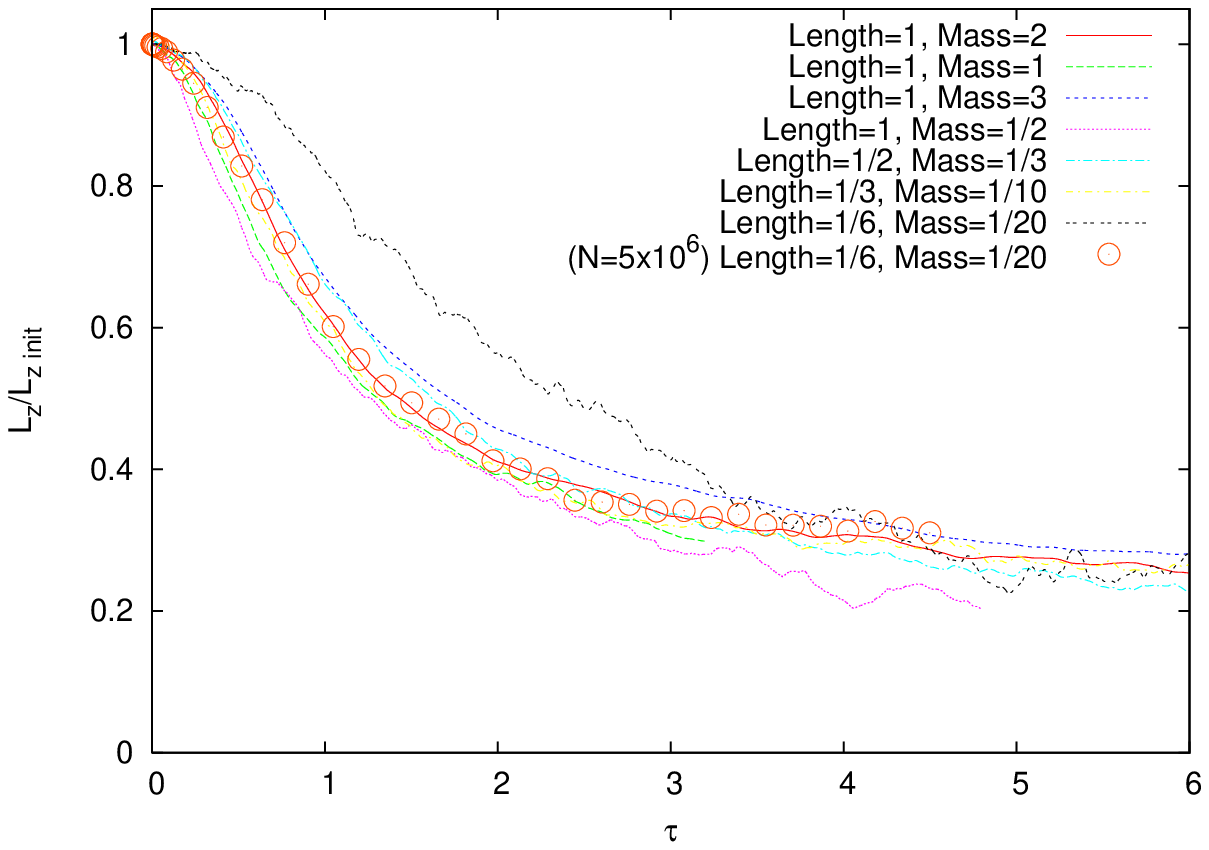}
  }
  \subfigure[50\% fast-limit + 50\% slow-limit scaling]{
    \includegraphics[width=0.49\linewidth]{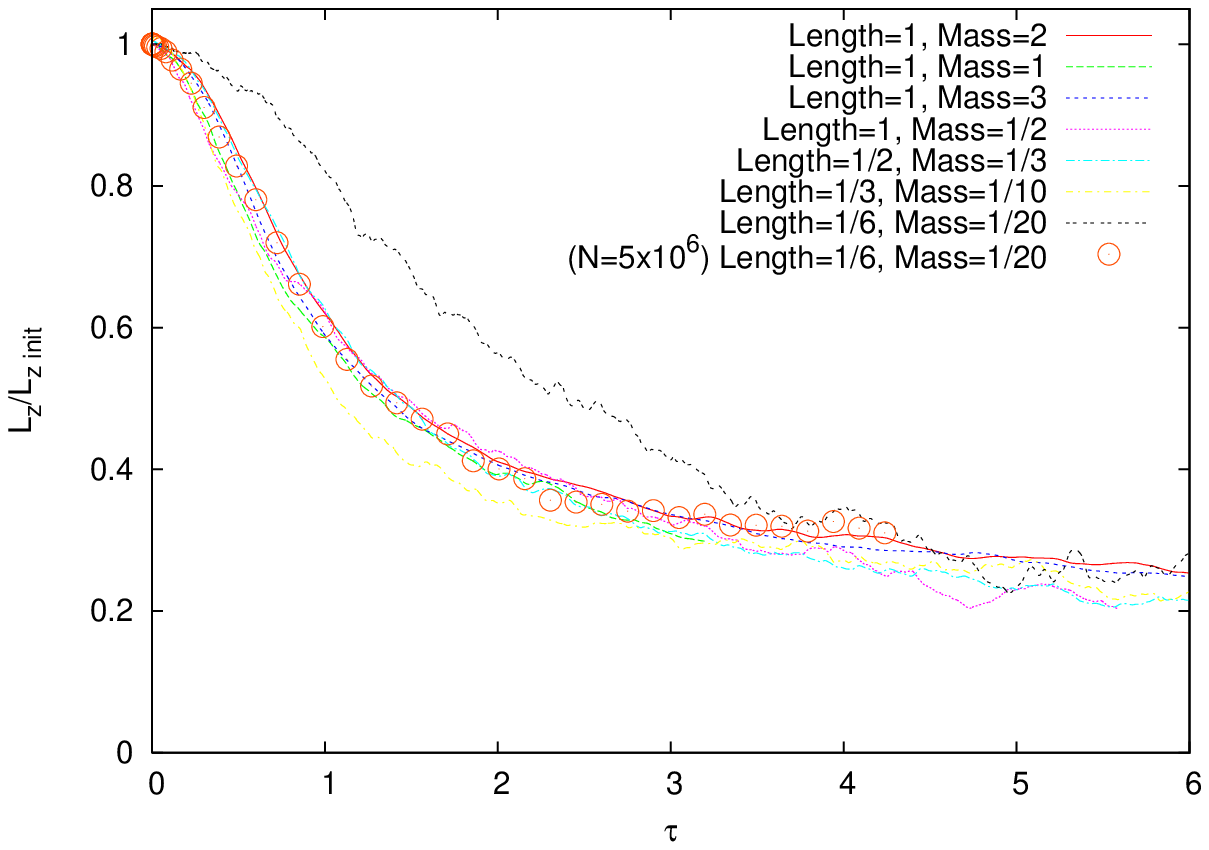}
  }
  \caption{The change in $L_z$ for slowing bars of different
masses and sizes scaled to the fiducial run labelled as Length=1,
Mass=1. The circles represent the evolution of the Length=1/6 case
using 5 times more particles.}
  \label{fig:scaled}
\end{figure}

In the past section we discussed the results for our fiducial bar,
which is much larger than bars found in galaxies today.  Here, we
discuss the evolution of both shorter bars and those with different
masses.  The different bar lengths and masses that we investigate are
presented in Table 1.  Since the inner parts of a NFW profile are a
power law, it is scale free inside of $r_s$, which is the length of
our largest bar.  Therefore, the evolution of shorter bars should
scale as the bar remains the same fraction of the enclosed mass, and
the corotation radius remains the same number of bar lengths.
Perturbation theory predicts a well-defined scaling of the overall
angular momentum evolution with bar mass, more precisely with the
perturbation strength. However, the scaling depends on the nature of
the resonant interaction.  If the evolution is fast enough, an orbit
passes through the resonance quickly compared to the period of the
resonant orbit.  As long as the angular momentum is transferred in
this {\em fast} limit \citep[hereafter TW]{Tremaine.Weinberg:84}, the
torque scales as $M_b^2$.  We saw in Paper I that some resonances and
especially ILR have contributions from the {\em slow} limit, for which
the torque scales as $M_b^{1/2}$.  Moreover, for strong bars, most of
the low-order resonances are in the transition region between these
two regimes, rapid slowing not withstanding.  The interactions that
lead to orbits exchanging angular momentum at resonance are described
in Paper I.  The total angular momentum of a bar is proportional to
its mass:
\begin{equation}
L_z = I_b\Omega_p \propto M_b R_b^2\Omega_p
\end{equation}
If we assume that all angular momentum transfer is in the fast limit, then
\begin{equation}
  {\dot L}_z \propto M_b^2
\end{equation}
and we can compare simulations of different bar mass $M_b$, bar length
a, and bar growth time $A(t)$ as follows. Let the ratio of the bar
mass to the halo mass inside of the bar radius be ${\cal M}\equiv
M_b/M(a)$.  If we scale the time by ${\cal M}/\Omega_p$ and the
angular momentum, $L_z$, by $L_{z\,init} {\cal M}$ then all the
evolution histories should be the same. Remember that this is if the
fast-limit scaling applies.  Alternatively, if we assume that all the
angular momentum transfer is in the slow limit, then
\begin{equation}
  {\dot L}_z \propto M_b^{1/2}.
\end{equation}
and We must scale the time by ${\cal M}^{-1/2}/\Omega_p$.  Finally, we
must take into account the bar growth time, $A(t)$.  Since the torque
mechanism is secular, we expect the degree of evolution to be
proportional to the time integral of the applied amplitude, $\tau =
\int_0^t dt\, A(t)$ as defined in \S\ref{sec:fid}.

\begin{table}
  \caption{Bar parameters}
  \begin{tabular}{llll}
    Length	& Mass	& Real length ($a$)	& $M_b/M_{halo}(a)$ \\ \hline
    1		& 1/2	& 0.067			& 0.25	\\
    1		& 1	& 0.067			& 0.5	\\
    1		& 2	& 0.067			& 1.0	\\
    1		& 3	& 0.067			& 1.5	\\
    1/2		& 1/3	& 0.033			& 1.0	\\
    1/3		& 1/10	& 0.02			& 0.5	\\
    1/6		& 1/20	& 0.01			& 0.5	\\ \hline
  \end{tabular}
  \label{tab:barparm}
\end{table}

Figure \ref{fig:scaled}a shows the evolution of $L_z$ with $\tau$ for
simulations with different bar lengths and masses, labelled in units
of the fiducial model, as outlined in Table 1, and scaled for the fast
limit.  These simulations also use $N=10^6$ multimass particles.  Bar
corotation occurs at $a$ for $t=\tau=0$ in all of these simulations.
The first four cases in the Figure show the scaled pattern speed
evolution for the Length=1 bar with a range of masses. In the last
three cases we reduce both the bar length and the bar mass.  The
smallest bar, Length=1/6, is approximately a disk scale length when
scaled the Milky Way.  With the rescaling the evolutions look similar,
except for that of the Length=1/6, Mass=1/20 bar, which we discuss
more below.  The residual differences could be caused either by
non-linear interactions, which could cause a breakdown in the linear
scaling, numerical artifacts, or contributions by resonances in the
slow limit.  We make a rough test of the last possibility by taking a
equal-part linear combination of the fast-limit and slow limit
scalings in Figure \ref{fig:scaled}b.  Now all but two of the cases
are almost identical; the two discrepant ones are the Length=1/3,
Mass=1/10 and Length=1/6, Mass=1/20 bars. Since these two cases are
for the least massive bars it seems unlikely that their inappropriate
scaling is related to non-linear effects.

Even with these differences, the scaling of $L_z(t)$ is remarkably
good over an order of magnitude in unscaled time and, therefore, many
of these bars have very different rates of slowing.  This may be
understood as follows.  If we think of performing a periodogram on the
bar perturbation, we will see a broadened line in frequency space; the
broadening will be inversely proportional the change in $\Delta
L_z/L_z$ since the amount of torque lost by the bar controls its
instantaneous frequency.  The finite time since the formation of the
bar also contributes to the broadening.  However, the total power,
determined by integrating under the broadened line, will be
approximately the same since the total rotational kinetic energy is
approximately the same.  For most resonances, except ILR, the
resonance occupies a narrow band in energy.  However, for ILR, the
energy range may be large.  Rapid evolution and subsequent broadening
in frequency space can further increase the importance of the ILR
contribution \citep[Paper I]{Weinberg:04}.

\begin{figure}
  \includegraphics[width=\figscaleA\linewidth]{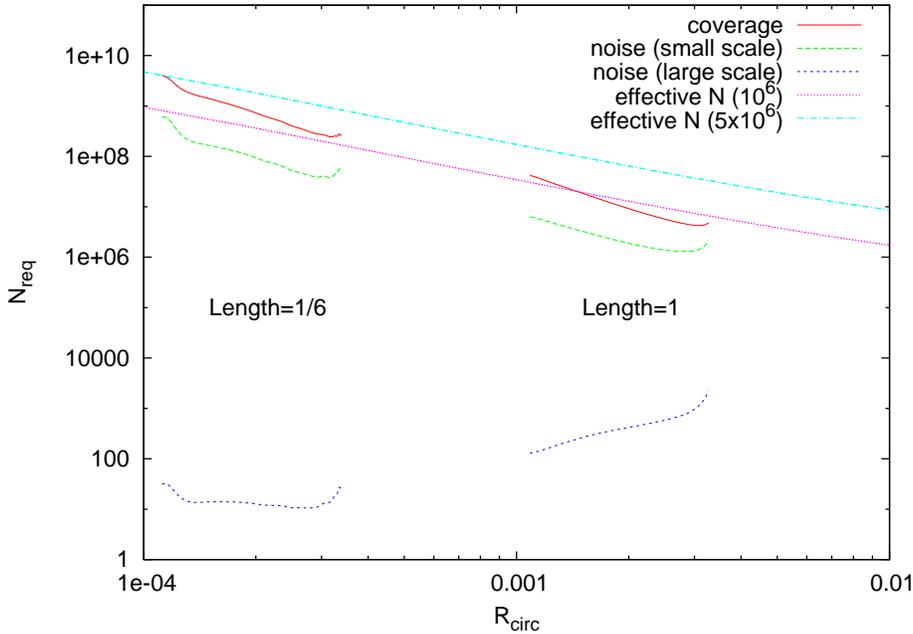}
  \caption{Different particle number criteria from Paper I for a
    scale-length bar (Length=1/6) and the fiducial bar (Length=1) for the
    ILR resonance in terms of equal mass particles within the virial radius.
    Also shown are the effective number of particles for our multimass 
    simulations using $10^6$ and $5\times 10^6$ particles.}
  \label{fig:innernreq}
\end{figure}

The scaling failure of the small length bars evolutionary histories is
most likely caused by numerical deficiencies. In Paper I, we derived
three requirements on particle number to model resonant dynamics
correctly within N-body simulations.  We present these minimum
particle number requirements for both the fiducial, Length=1 bar
(right-hand side) and the scale-length, Length-1/6 bar (left-hand side)
in Figure \ref{fig:innernreq} as a function of radius.  Each resonance
has different requirements and we show those for the resonance that is
most important for both bar slowing and dark matter cusp evolution,
the ILR. We only plot the criteria over the range in radii that
dominate the ILR for each bar length.  Since we plot the number of
equal mass particles required within the virial radius, to compare
with our multimass simulations, we also plot the equivalent number of
equal mass particles in our simulations. Note that in our SCF code,
which has no direct two-body interactions, the small scale noise
criterion does not apply.  Nevertheless, for both bars the coverage
criterion is the most stringent.  For the small, Length=1/6 bar, one
requires about $10^9$ particles over the range of radii that dominate
the ILR.  The same coverage criterion demands only about $10^7$
particles for the larger fiducial bar.  Comparing this with the
effective number of particles in our $10^6$ multimass simulations, one
can see that the fiducial length bar has a sufficient number of
particles to follow the resonant dynamics and the small, Length=1/6
bar simulation does not, as is born out in its poor scaling.  To test
this idea further, we repeated the Length=1/6 bar simulation using
five times more particles, i.e. $5\times 10^6$ multimass particles,
which according to Figure \ref{fig:innernreq} should be sufficient.
Now the scaled evolutionary history follows that of the other
simulations, plotted as the open circles in Figure \ref{fig:scaled}.
This demonstrates the extreme sensitivity of the evolution to
numerical deficiencies and the need for a detailed understanding of
the dynamics in order to trust the results of N-body simulations.

\begin{figure}
  \subfigure[Length=1/2]{%
    \includegraphics[width=0.5\linewidth]{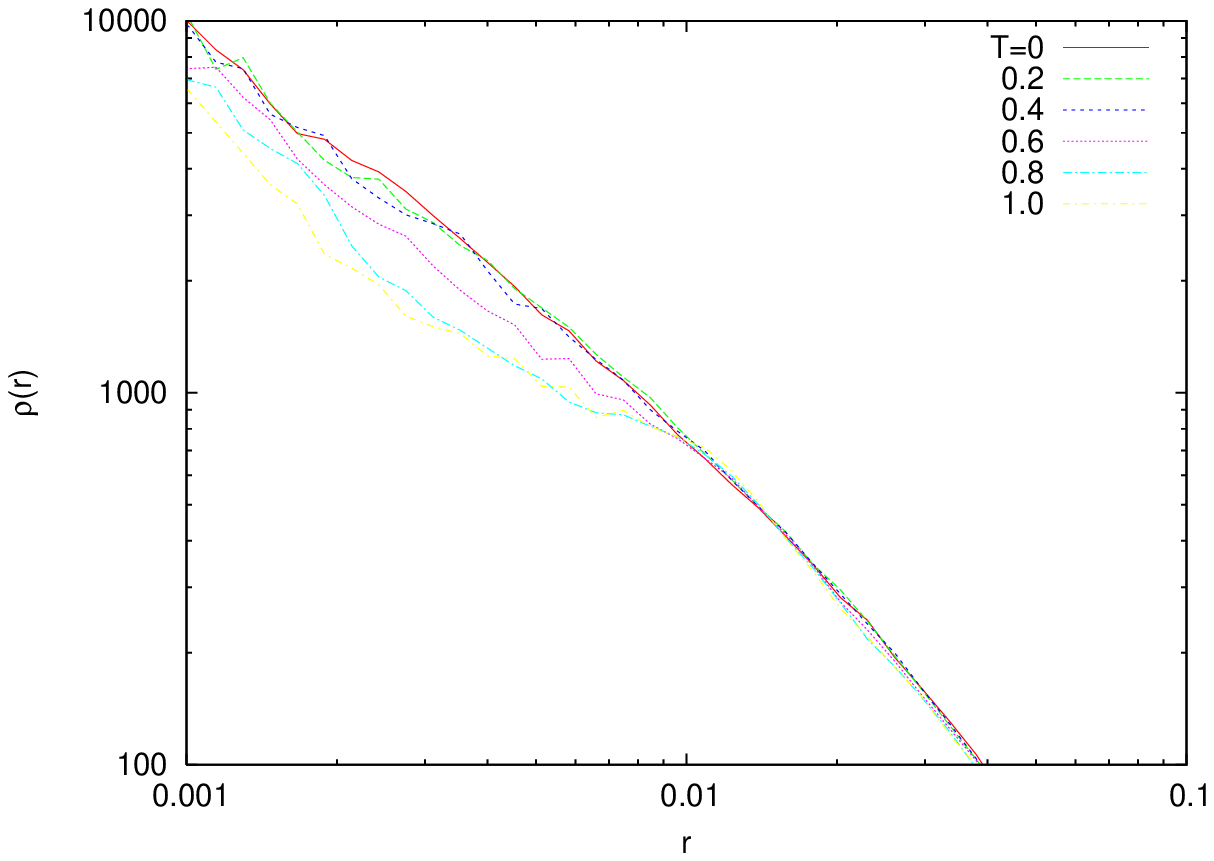}
  }
  \subfigure[Length=1/3]{%
    \includegraphics[width=0.5\linewidth]{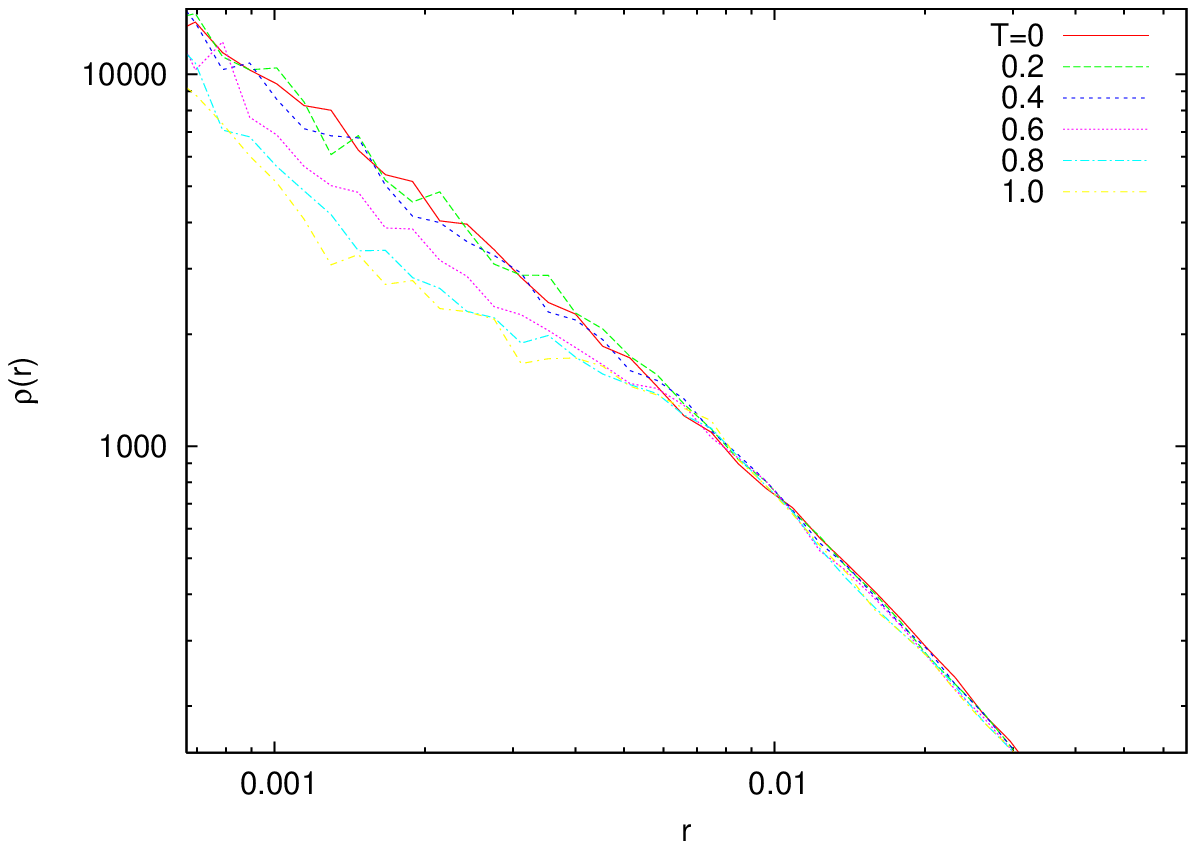}
  }
  \subfigure[Length=1/6]{%
    \includegraphics[width=0.5\linewidth]{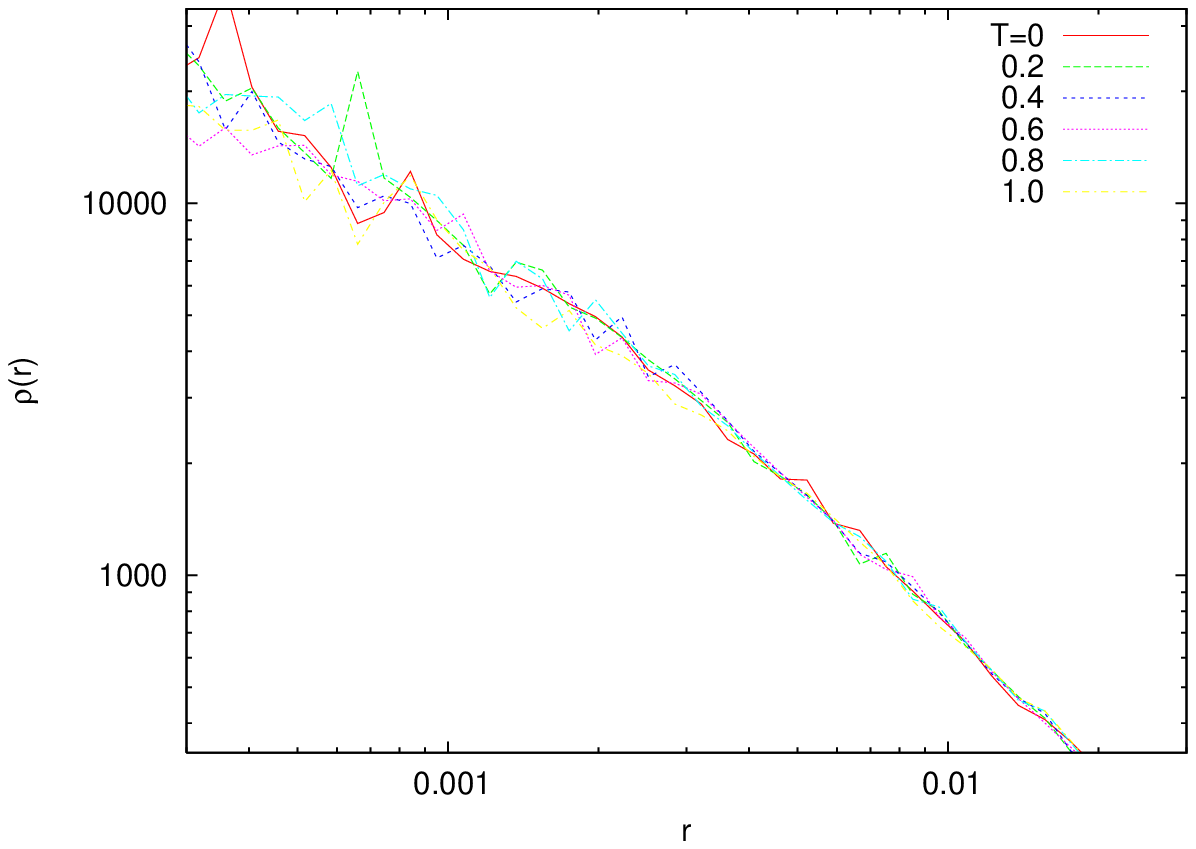}
  }
  \subfigure[Length=1/6, $N=5\times10^6$']{%
    \includegraphics[width=0.5\linewidth]{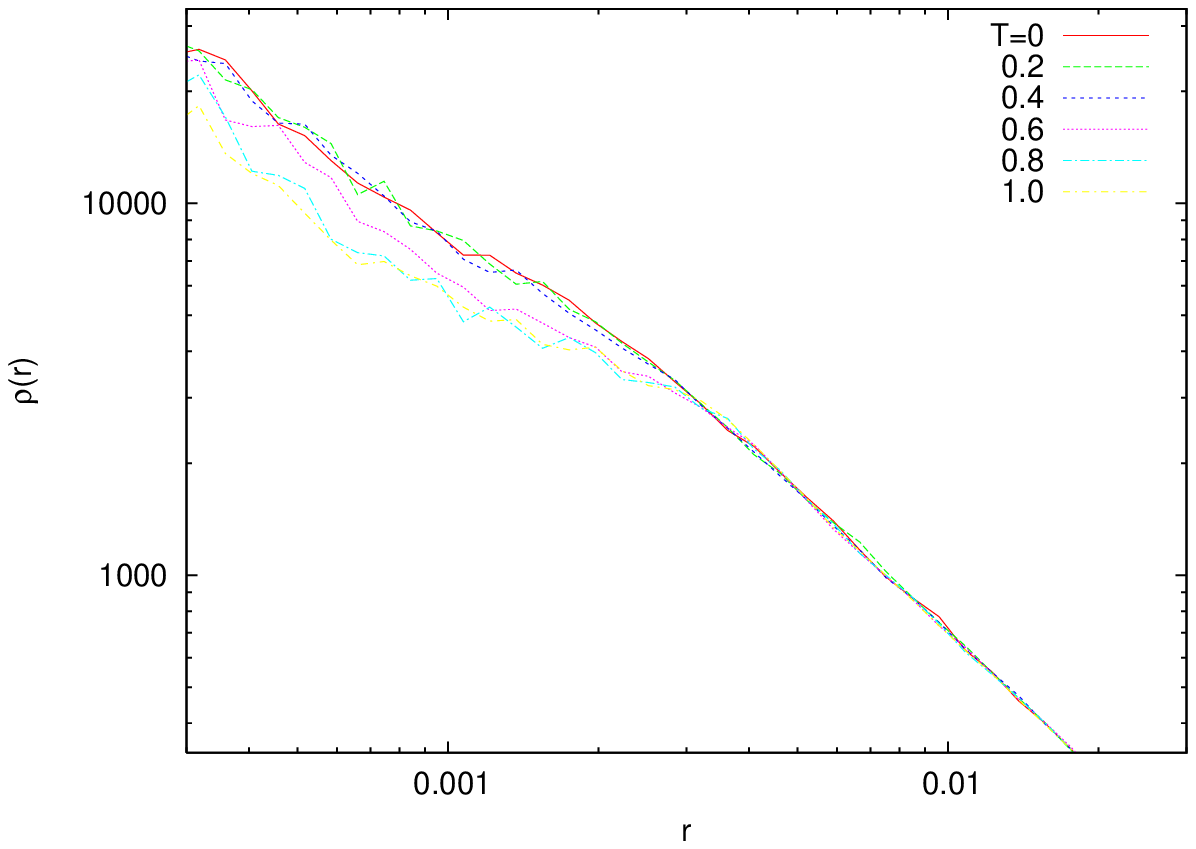}
  }
  \caption{The density profile evolution for three shorter bars.
    Simulations in the first three panels
    use $N=10^6$ multimass particles and the fourth uses
    $N=5\times10^6$ multimass particles.}
  \label{fig:scaledprof}
\end{figure}

Like the longer fiducial bar, the shorter bars, except for the
Length=1/6 bar, also cause density profile evolution, as shown in
Figure \ref{fig:scaledprof}abc.  For the Length=1/2 ($a=0.033$ or
three disk scale lengths) bar and Length=1/3 ($a=0.02$ or two disk
scale lengths) bar one clearly sees the same profile evolution as for
the fiducial bar in \S\ref{sec:fid}.  As predicted by Figure
\ref{fig:innernreq}, there is no profile evolution in the Length=1/6
($a=0.033$ or one disk scale length) bar simulation with $10^6$
multimass particles.  However, once again if we use $5\times 10^6$
multimass particles, a sufficient number to resolve the ILR resonance
responsible for the cusp density evolution according to Figure
\ref{fig:innernreq}, density profile evolution occurs just as in the
three longer bars as shown in Figure \ref{fig:scaledprof}d.  For all
four bar lengths the density is reduced within about 30\% of the bar
radius, independent of bar length if the numerical criteria are
satisfied.

In summary, these examples also help clarify several important
features of bar evolution in the simulations.  For an astronomically
realistic bar ending at corotation, a naive estimate places the ILR
deep within the cusp. For a scale length bar, the particle number
predictions from Paper I suggest that the ILR should not be seen in
our simulations with $10^6$ multimass particles, and therefore there
should be no density profile evolution and indeed there is not.
However, for $5\times10^6$ particles, the effective number of
particles near ILR exceeds the coverage criterion and we again see
density evolution.  Even so, the profiles in the simulations do not
evolve at very small radii.  It is now clear that these simulations
cannot couple to these small radii solely for computational reasons.
Finally, although the multimass initial conditions used here improved
the radial resolution and allowed us to recover the correct resonant
dynamics using fewer total particles, this technique must be
cautiously because it can produce more noise at larger radii.

\subsection{Variation in bar shape}
\label{sec:barshape}

\begin{figure}
  \centering
  \subfigure[$a_2/a_1=0.5$]{\includegraphics[width=0.45\linewidth]{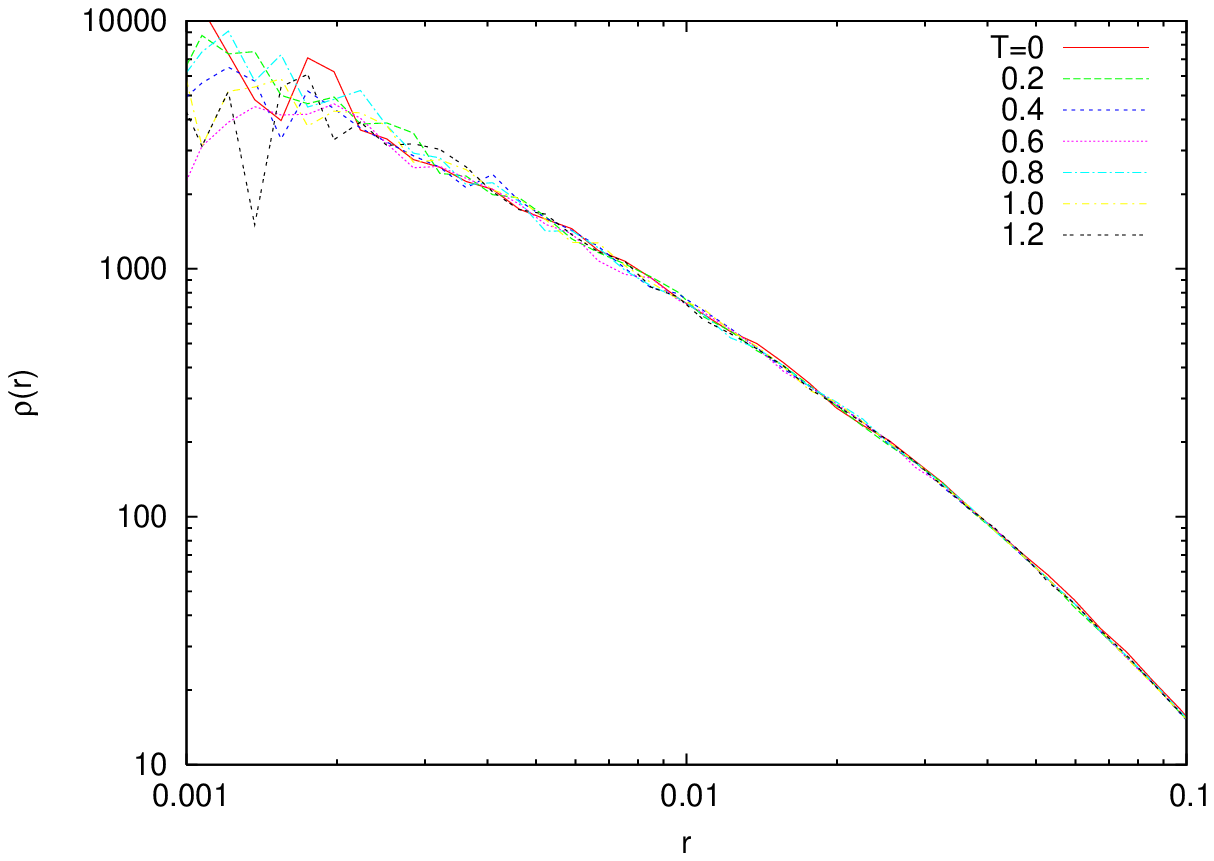}}
  \subfigure[$a_2/a_1=0.3$]{\includegraphics[width=0.45\linewidth]{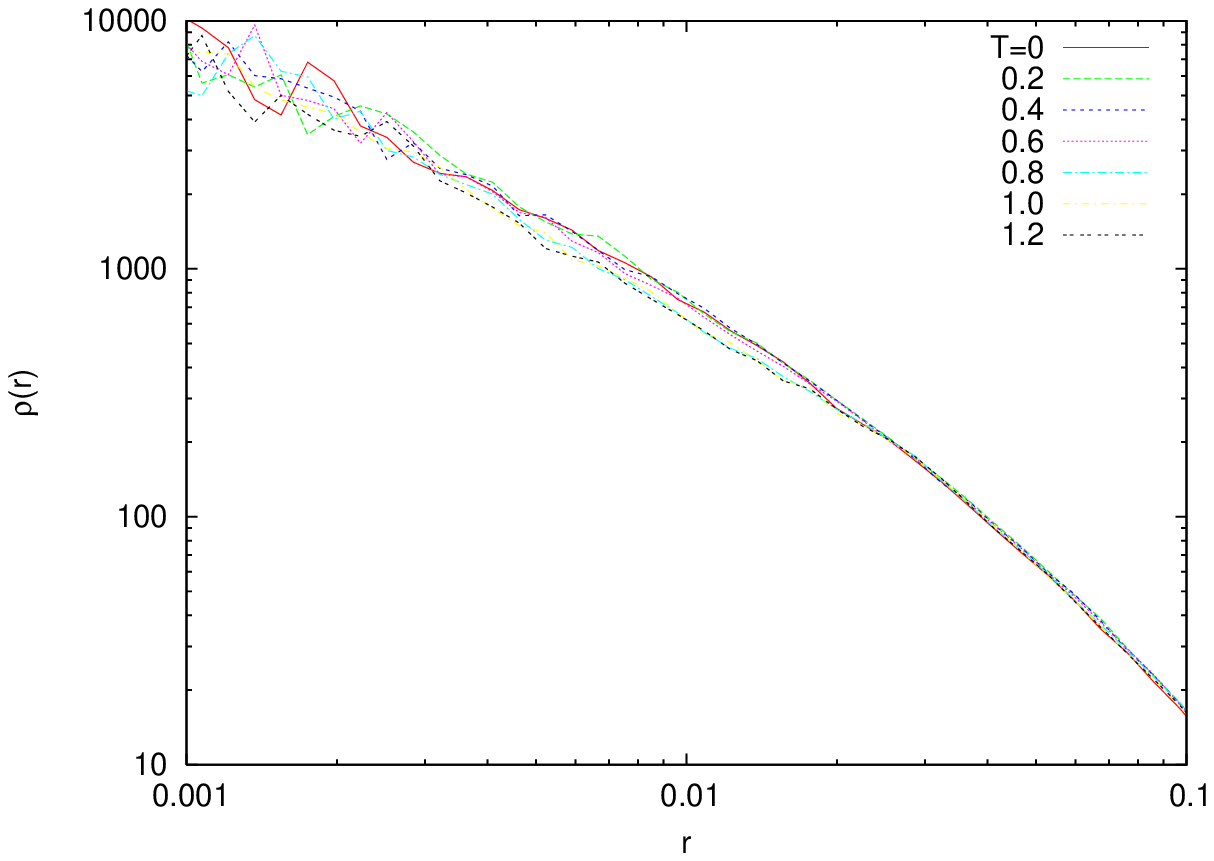}}
  \subfigure[$a_2/a_1=0.2$]{\includegraphics[width=0.45\linewidth]{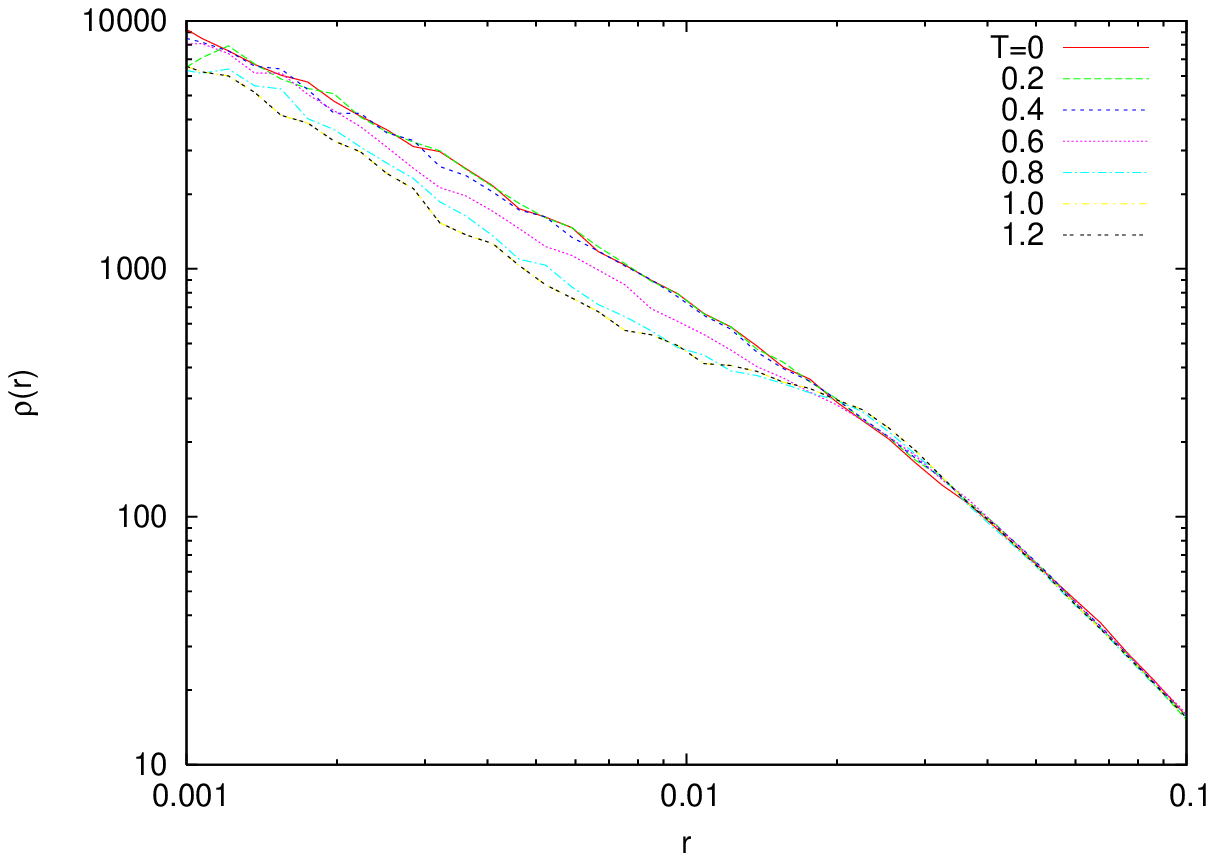}}
  \subfigure[$a_2/a_1=0.15$]{\includegraphics[width=0.45\linewidth]{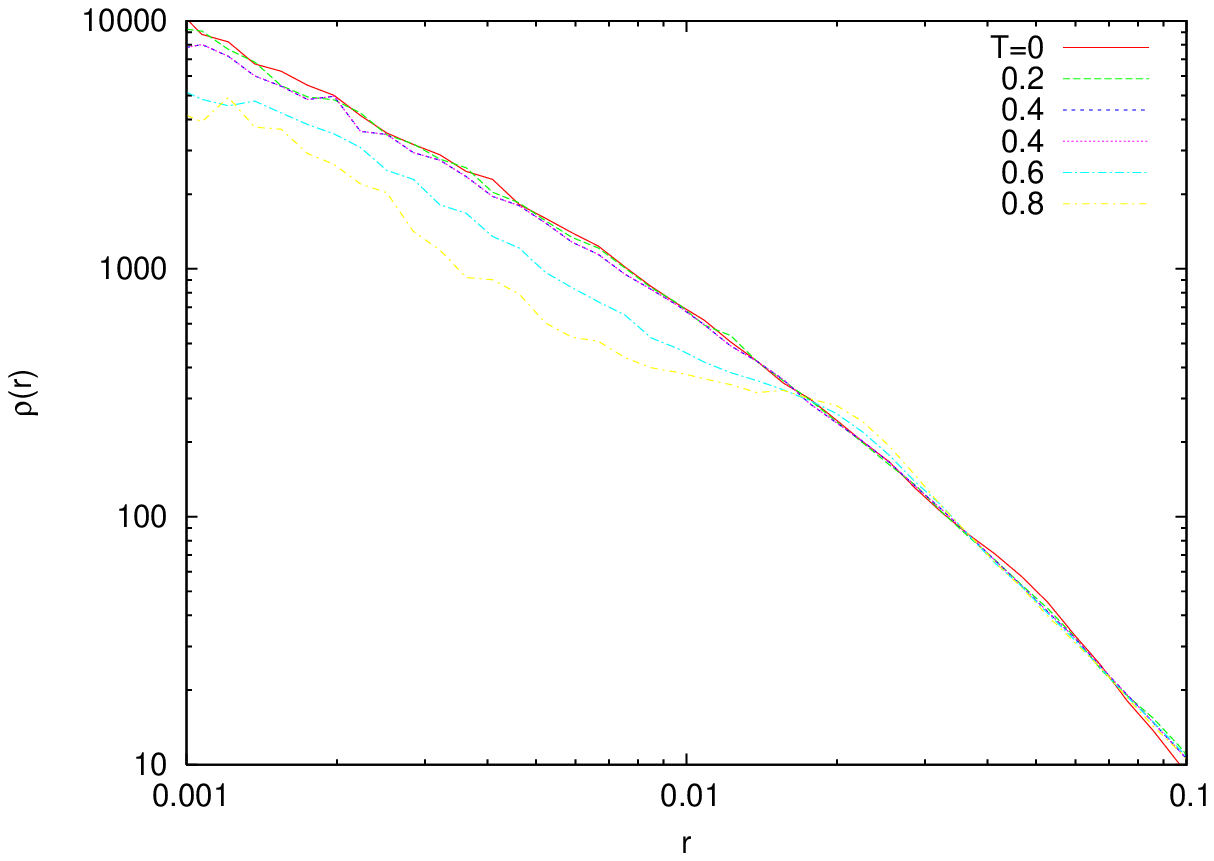}}
  \caption{Profile evolution with bars of different axis ratio. Each
    curve is labelled by time.  The initial bar rotation period is 0.3
    time units.}
  \label{fig:axisrat}
\end{figure}

Figure \ref{fig:quadcomp} clearly shows that a larger major to
semi-major axis ratio results in a smaller quadrupole amplitude inside
of the bar radius.  However, if the quadrupole amplitude becomes too
small, there will no way of coupling to the inner halo.  We present a
series of simulations with $N=10^6$ (without multimass) with an
evolving bar pattern speed to investigate this anticipated trend.  The
evolution of the halo profiles for different axis ratios is presented
in Figure \ref{fig:axisrat}.  Otherwise, parameters are those from the
fiducial run.  Each panel shows a different axis ratio $a_2/a_1$ with
fixed $a_3/a_1$. There is little evolution in the profile for
$a_2/a_1\ga0.3$ there is no obvious evolution.  However, there is
clear inner profile evolution for $a_2/a_1\la0.2$.

\begin{figure}
  \centering
  \subfigure[small-scale noise]{%
    \includegraphics[width=0.45\linewidth]{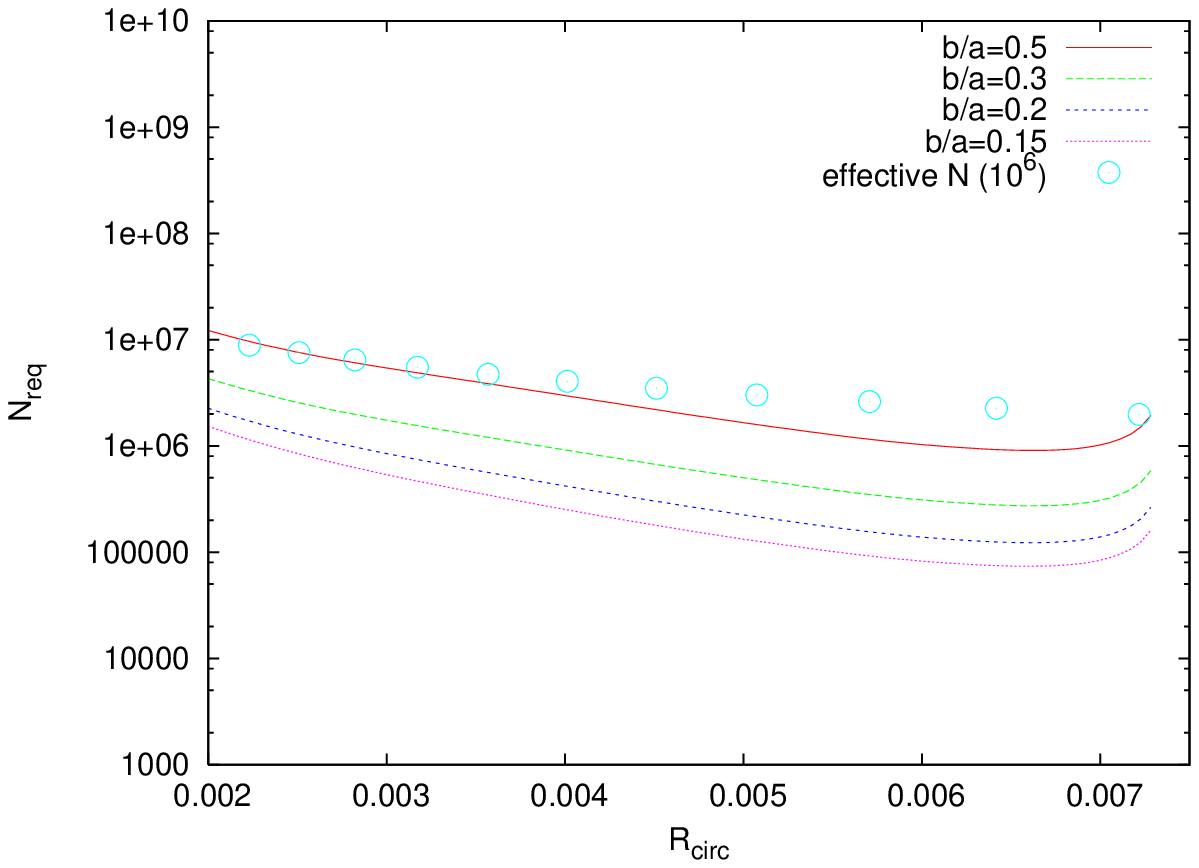}
  }
  \subfigure[large-scale noise]{%
    \includegraphics[width=0.45\linewidth]{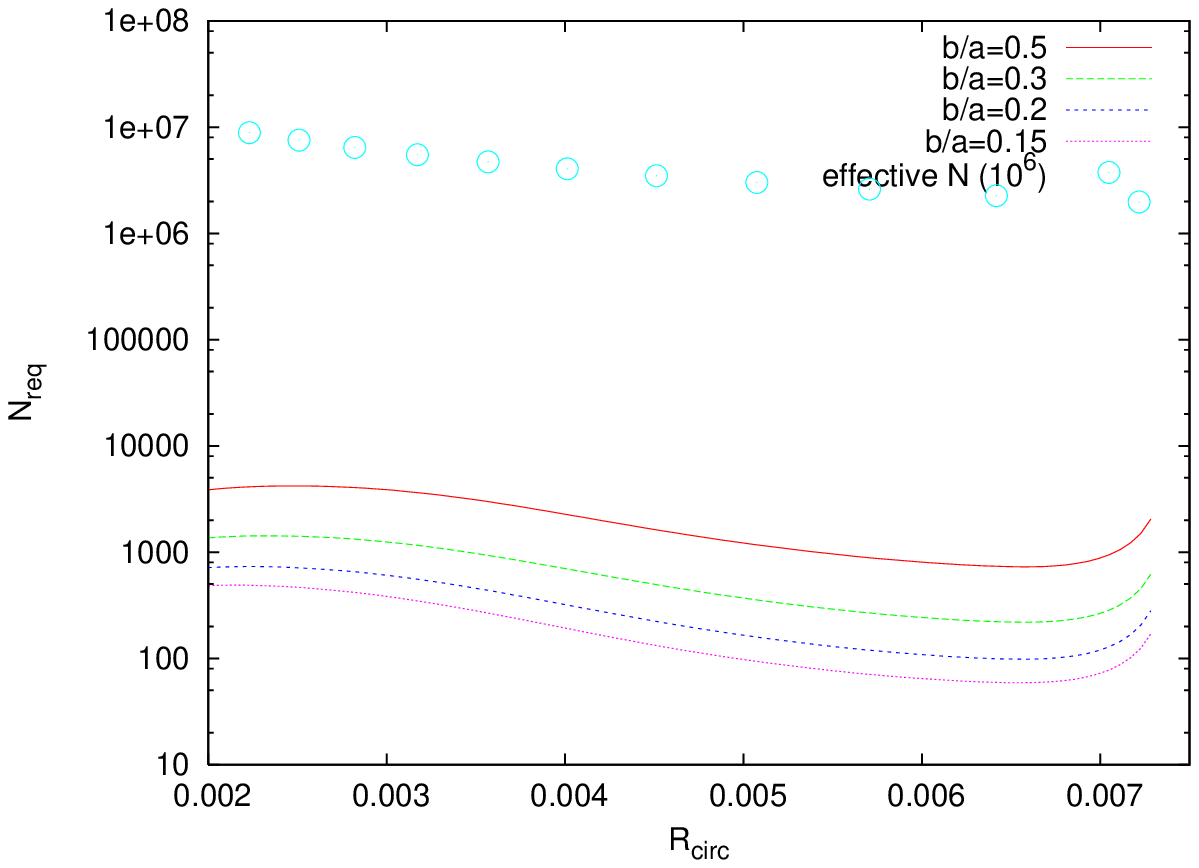}
  }
  \subfigure[coverage]{%
    \includegraphics[width=0.45\linewidth]{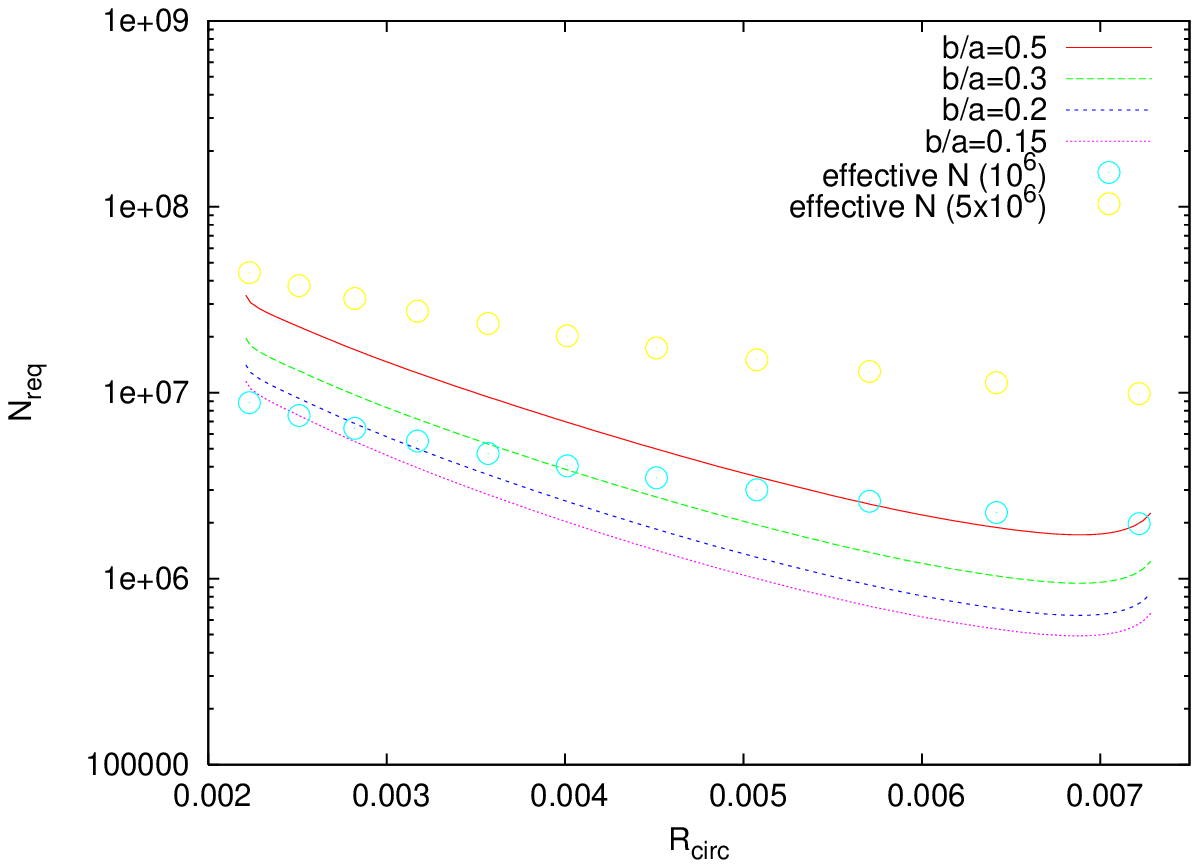}
  }
  \caption{Critical particle number criteria from Paper I for
    fiducial bar ($a=0.067$) and but with varying semi-minor to
    semi-major axis ratios.}
  \label{fig:resolve}
\end{figure}

This behaviour is expected from on the criteria described in Paper I
(see Fig. \ref{fig:resolve}).  The coverage criterion is dominant here
as discussed earlier. The sharp threshold is the combined result of
increasing amplitude with decreasing axis ratio and more rapid slowing
with increasing amplitude.  A smaller amplitude perturbation in the
vicinity of a resonance yields a smaller resonance width and therefore
a higher central particle number requirement.  Similarly, a weakly
slowing bar will not broaden the ILR to lower frequencies and
therefore larger radii where the coverage criterion is also less
stringent.  These runs suggest that our fiducial simulation is close
to the minimum necessary particle numbers.  Changing $a_2/a_1$ from
0.2 to 0.5 makes an order of magnitude decrease in the value of the
inner potential.  It is, therefore, no surprise that we see critical
behaviour as we increase the amplitude of the quadrupole by decreasing
the axis ratio $a_2/a_1$.  The description of these trends were
presented in WK.  The noise criterion for point particles is also
shown in Figure \ref{fig:resolve}.  For these cases, the small-scale
noise and coverage criteria have similar magnitude.  The noise
criteria scales as the amplitude of the perturbation potential while
the coverage criterion scales as the square root of this value; this
explains the difference in range with varying $a_2/a_1$.  The values
for the large-scale noise criterion, relevant to our simulation
method, are negligible here.

\subsection{Variation in halo density profile}
\label{sec:profiles}

\begin{figure}
  \includegraphics[width=\figscaleA\linewidth]{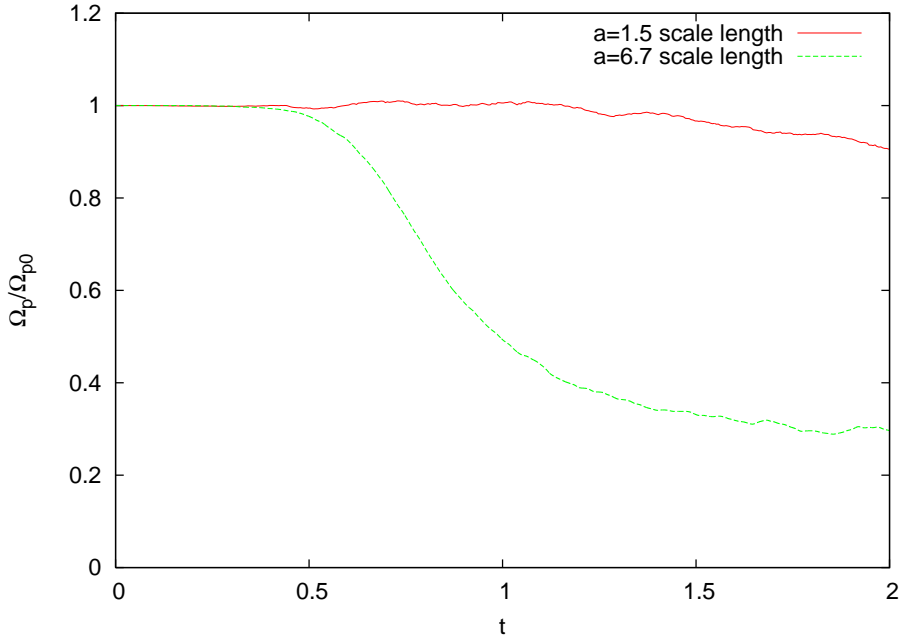}
  \caption{Pattern speed evolution in a King model with $r_c = r_s$ and
  two different bar models: the fiducial bar with semi major axis
  $a=0.067=r_s$ (6.7 disk scale lengths for the Milky Way) and a small
  bar with $a=0.015$ (1.5 disk scale lengths).}
  \label{fig:bigcore_comp}
\end{figure}

The shape of the halo density profile, which determines the existence
and location of resonances for a given bar pattern speed, therefore,
determines the magnitude and effect of the angular momentum transfer.
As an extreme example, we describe the evolution of a bar in a
lowered, truncated isothermal sphere \citep{King:66a} with a core
radius equal to $r_s$ in the NFW model, using the same total mass as
that within the virial radius in the fiducial run.  We consider two
bar lengths: one with a length of 0.015 (1.5 scale lengths when scaled
to the Milky Way) and one the same as the fiducial bar with a length
of 0.067 (6.7 scale lengths).  The mass of each bar is 50\% of the
mass of the enclosed dark matter halo mass and initially corotation
occurs at the end of the bar.  Resonant dynamic theory (see Paper I)
tells us that to have a resonance there must be orbits with the right
frequency.  In addition, within a constant density harmonic core, like
in the central regions of a King model, there is no resonance unless
$l_1 + l_2 - m =0$.  For the low order resonances, which are most
important for the evolution, this is only true for the corotation
resonance.  However, as the bar slows, the orbits with the right
frequency move further out in the halo and the corotation resonance
becomes very weak.  Therefore, there is no ILR in either case. In
addition, for the $a=0.015$ bar, there are also no (1,0) or (2,-2)
resonances, only a small region of OLR, and a weakening corotation
resonance.  Figure \ref{fig:bigcore_comp} shows the pattern speed
evolution for the two bars.  The difference in evolution of the
pattern speed, which owes to the different resonances present, is
striking.

\begin{figure}
  \subfigure[Profiles]{
    \includegraphics[width=0.50\linewidth]{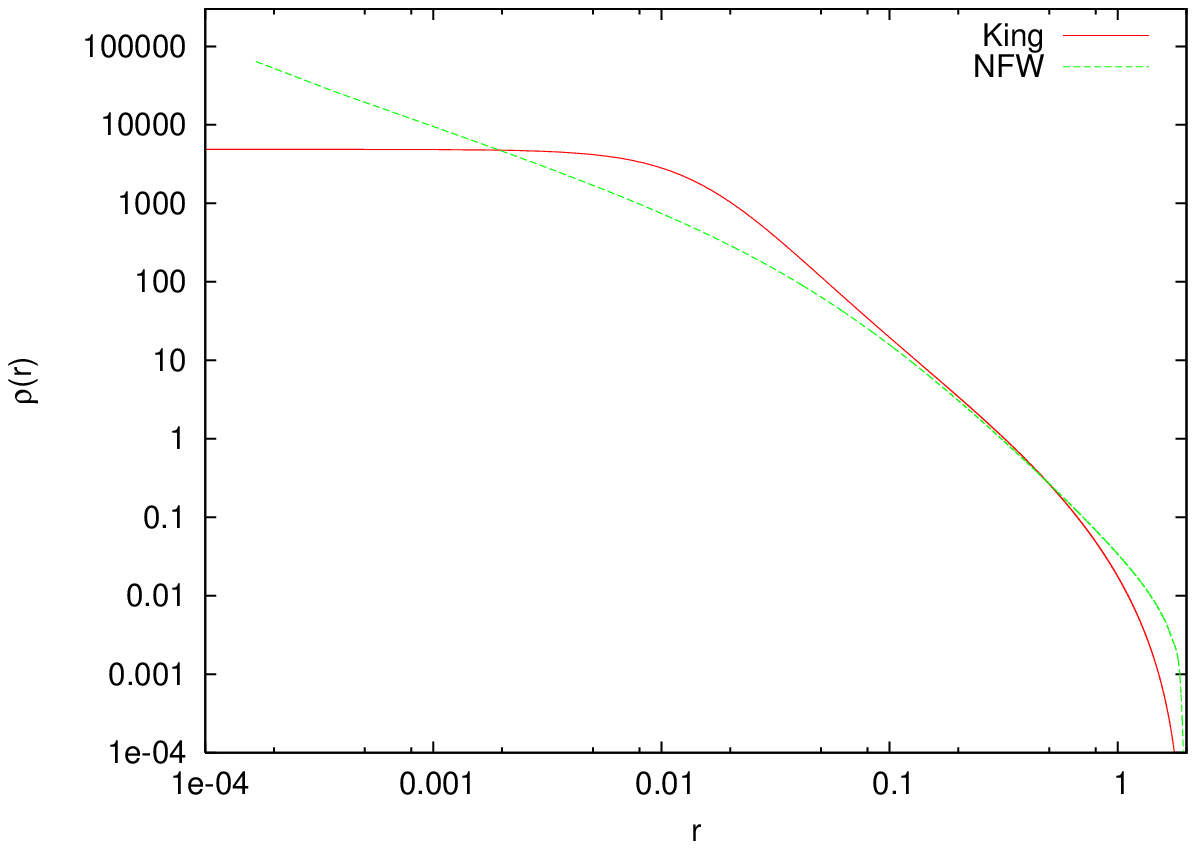}}
    \subfigure[Pattern speeds]{
    \includegraphics[width=0.50\linewidth]{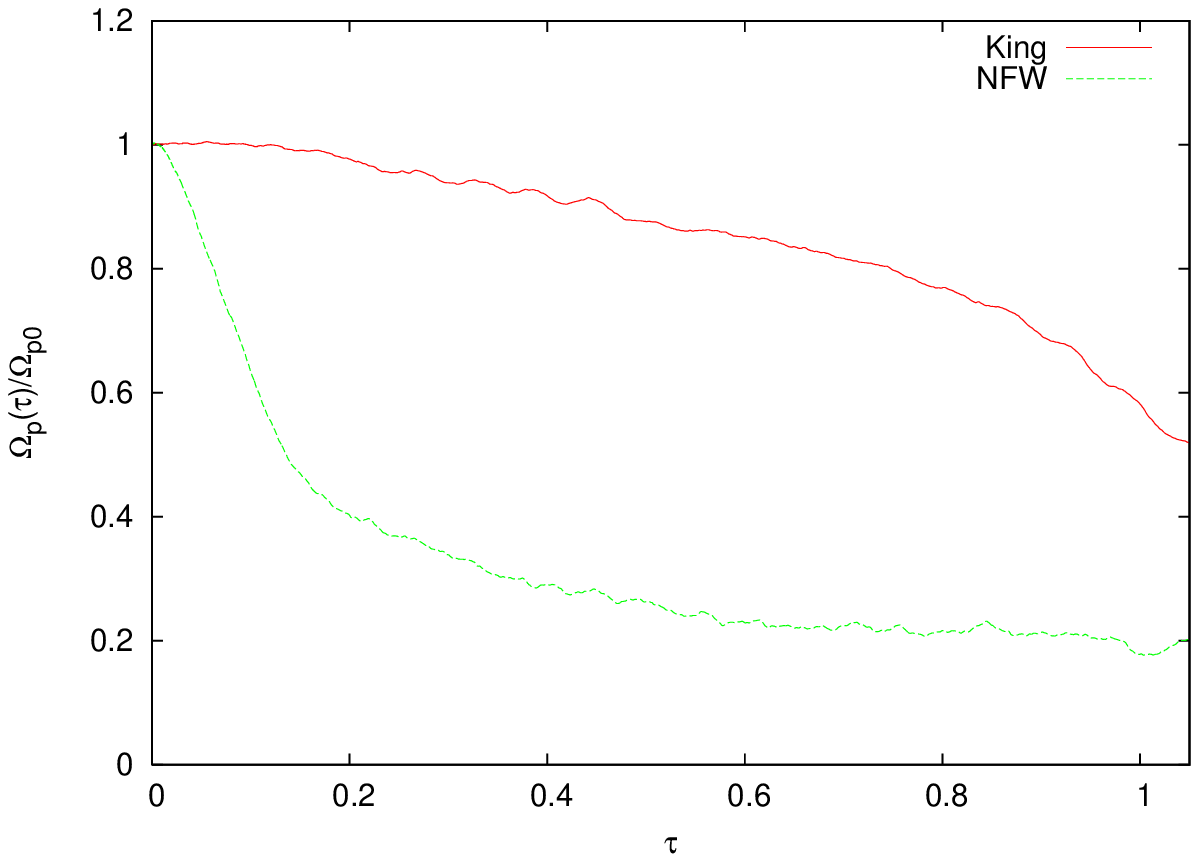}}
    \subfigure[$\Delta L_z$]{
    \includegraphics[width=\linewidth]{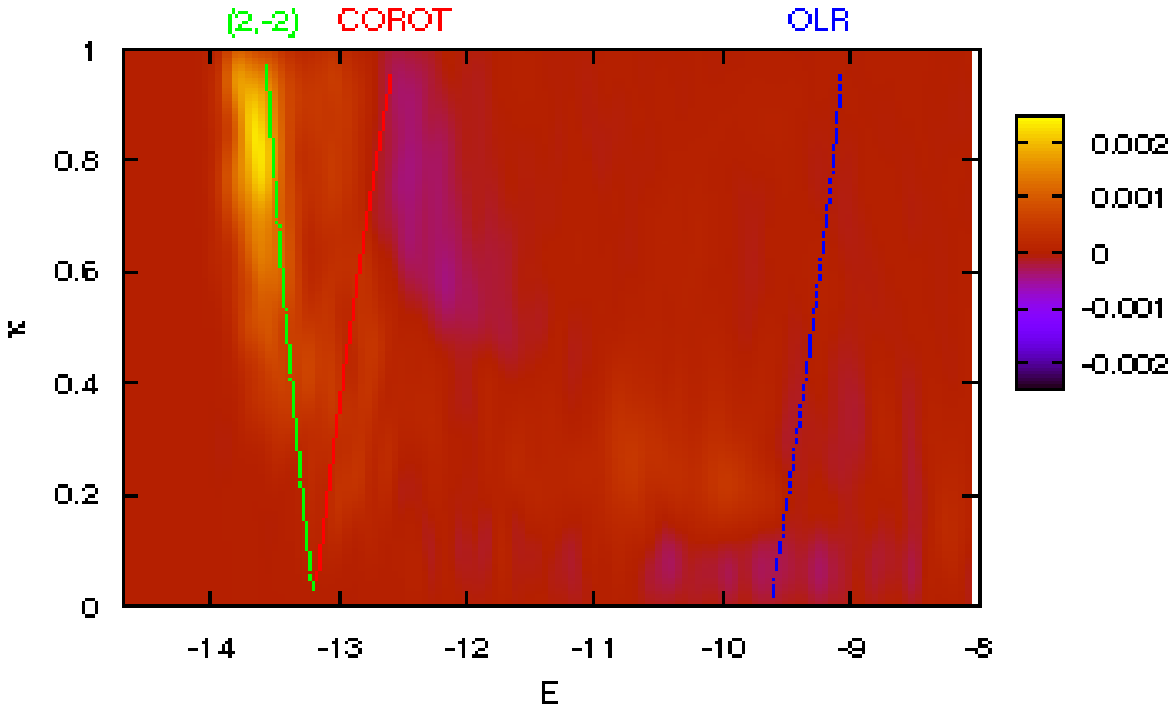}}
    \caption{Comparison of bar evolution in a King model dark halo
    with a NFW dark halo with similar concentrations and the same
    mass.  Panel (a) shows dark-matter density profiles, Panel
    (b)shows the evolution of the bar pattern speed and Panel (c)
    shows the distribution of angular momentum change in phase space
    for the King model.  We also overlay the location of the primary
    resonances as labelled.}
    \label{fig:king_nfw}
\end{figure} 

Next, we compare the evolution of a bar in a King model dark matter
halo with the evolution in a NFW dark matter halo.  We choose the King
Model to have the same mass and a similar concentration as the NFW
halo.  The bar has a length of 0.01 (one disk scale length when scaled
to the Milky Way) and initially corotation occurs at the end of the
bar.  Figure \ref{fig:king_nfw}a compares the two initial density
profiles.  The dark matter density of the King model at one bar length
is larger than that of the NFW model and the King density falls below
the NFW density at approximate 1/30 of the bar radius.  Figure
\ref{fig:king_nfw}b shows the evolution of the bar pattern speed. The
bar in the King model does not slow much at all because there is no
ILR or DRR, as shown in Figure \ref{fig:king_nfw}c where we show the
position of the resonances in phase space.  We have seen
(Fig. \ref{fig:dlz_compare}) that these two resonances are responsible
for a large fraction of the torque.  Although corotation, (2, -2), and
OLR are active, their net contributions nearly cancel.  Hence, these
resonances merely redistribute the angular momentum but do not provide
a significant \emph{net} torque on the bar.  This is an example of a
bar in massive dark matter halo that does not slow, although a
significant amount of total angular momentum is exchanged with the
halo. Often, people model bar slow down as Chandrasekhar dynamical
friction on two point masses, each with one half the bar mass, located
at the bar radius (e.g. Sellwood 2004).  In this case, such a
prescription would have predicted that the King Model bar would slow
faster than the NFW.  Local Chandrasekhar dynamical friction does not
apply to bar slow down, it is the result of resonant dynamics.

\begin{figure}
  \subfigure[Fiducial bar]{%
    \includegraphics[width=0.32\linewidth]{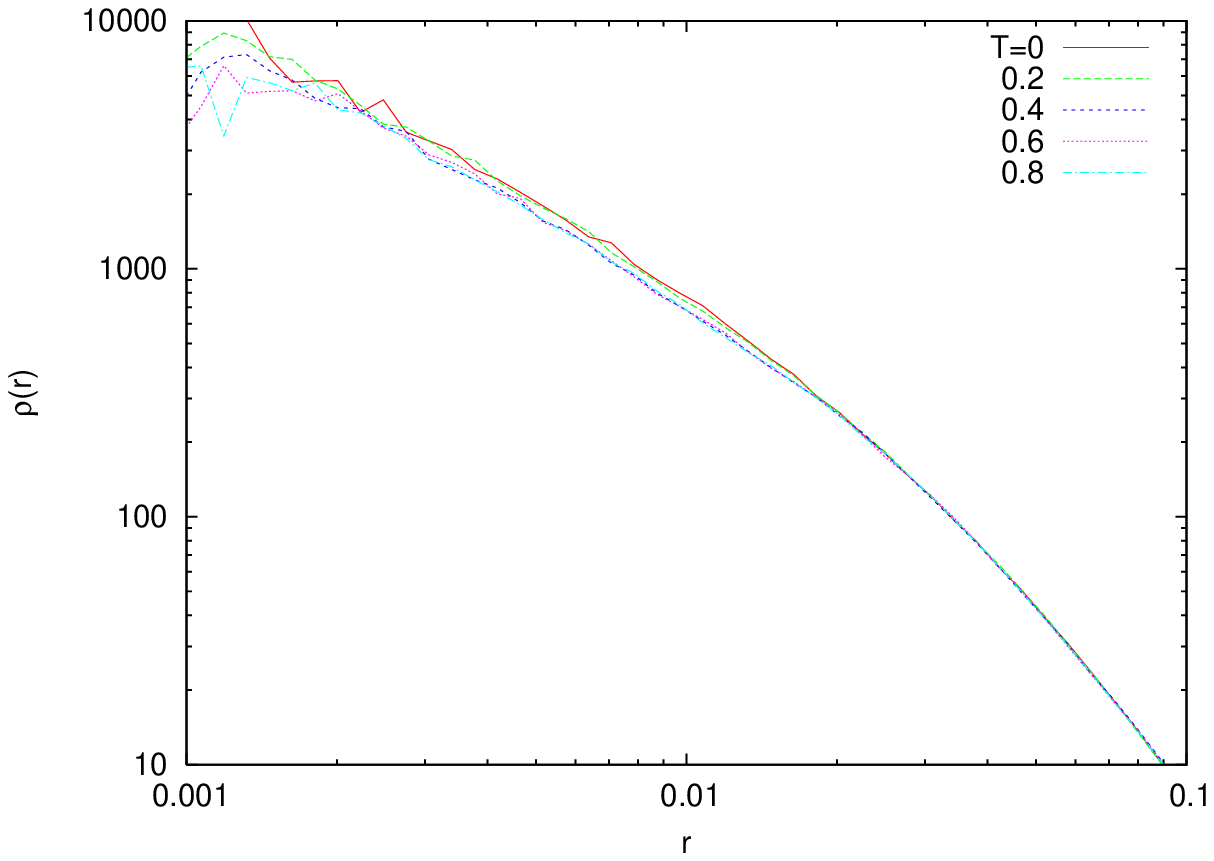}
  }
  \subfigure[2x mass]{%
    \includegraphics[width=0.32\linewidth]{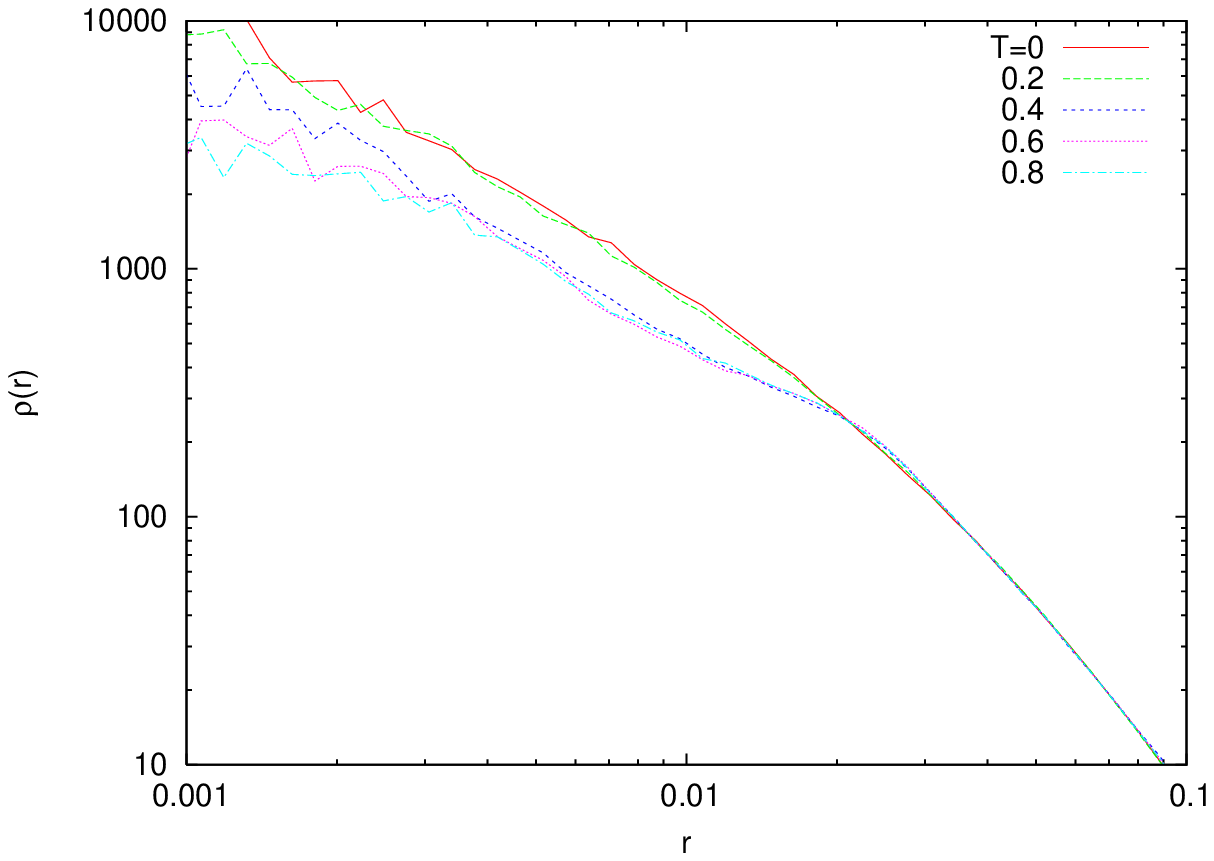}
  }
  \subfigure[2x mass, no self gravity]{%
    \includegraphics[width=0.32\linewidth]{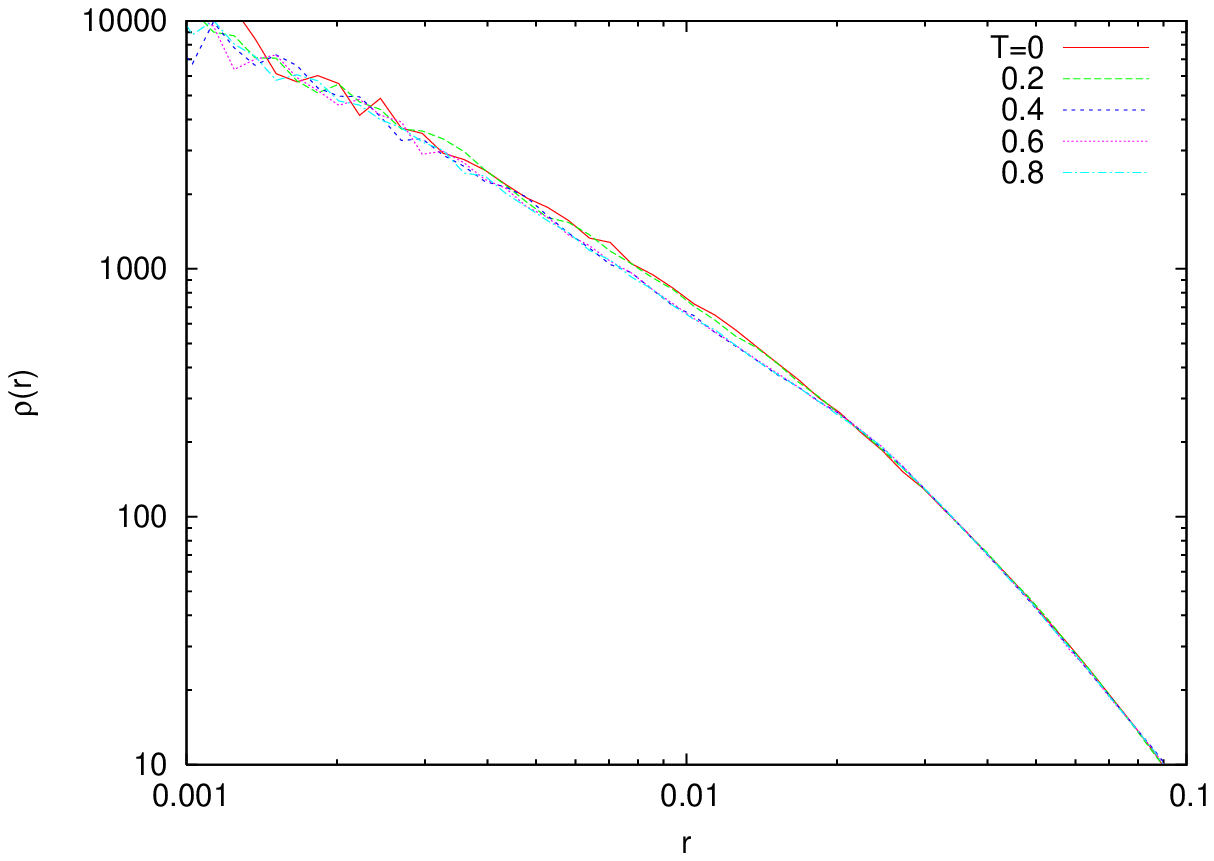}
  }
  \caption{Profile evolution in a Hernquist profile.  (a) The fiducial bar
    from \S\protect{\ref{sec:fid}} with $a=r_s$. (b) Twice the mass of
    the fiducial bar. (c) As in (b) but with no self gravity.}
  \label{fig:hern_fid}
\end{figure}

We have seen that all aspects of the galaxy model---all the bar and
dark matter halo parameters---can affect the subsequent evolution.
Therefore, when one compares the differences in evolution between
different scenarios, the values that are kept constant determine the
outcome.  For example, Figure \ref{fig:hern_fid}a shows the dark
matter halo evolution for our fiducial bar in a Hernquist dark matter
halo, $\rho\propto r^{-1}(r+a)^{-3}$, with $a=r_s$ and the same mass
within the virial radius as our NFW halo.  The profile evolution is
negligible.  However, the mass enclosed within the bar radius is 2.5
times larger than than that in the NFW model.  Increasing the mass of
the bar by a factor of two, which makes the bar mass 40\% of the
enclosed halo mass, leads to obvious profile evolution. For
comparison, Figure \ref{fig:hern_fid}c shows the evolution of the same
model \ref{fig:hern_fid}b but without self gravity.  The profile
evolves only slightly and demonstrates the importance of self gravity
in density profile evolution.

\begin{figure}
  \includegraphics[width=0.49\linewidth]{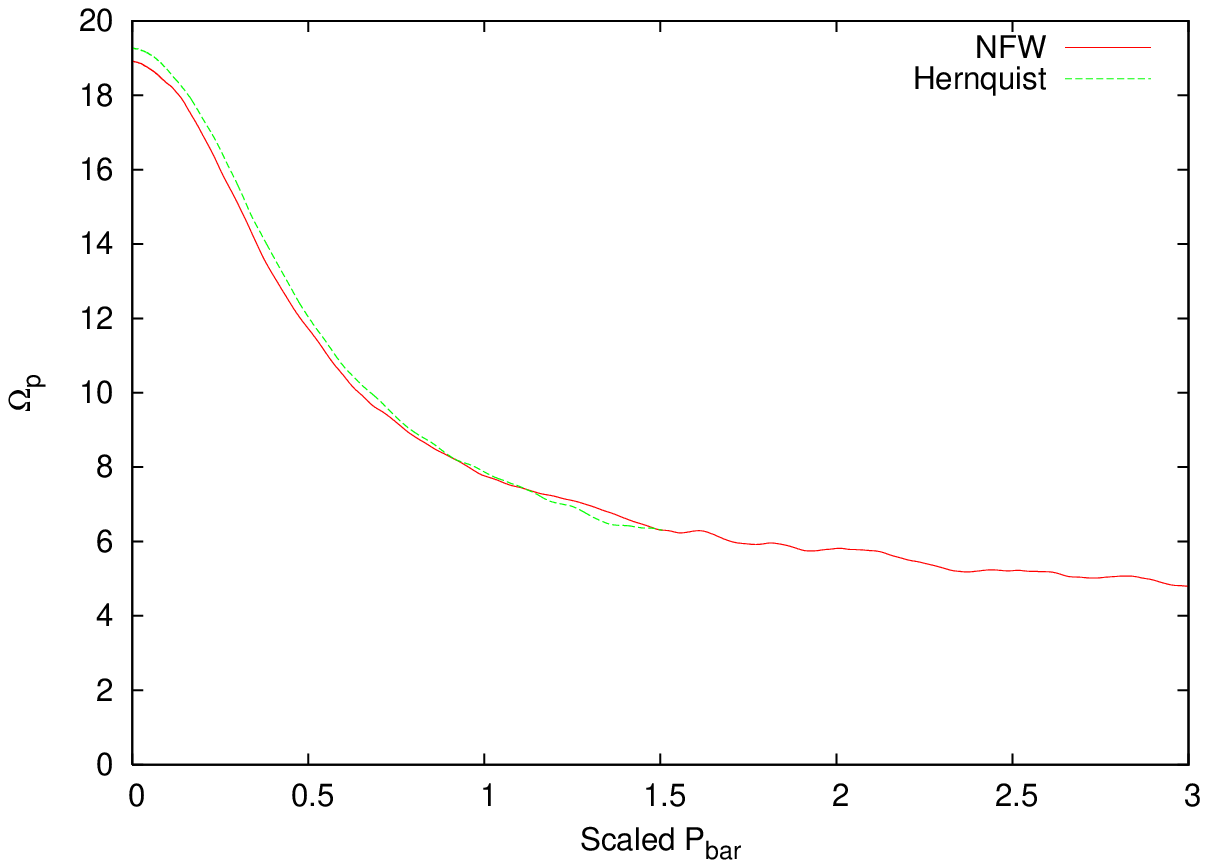}
  \includegraphics[width=0.49\linewidth]{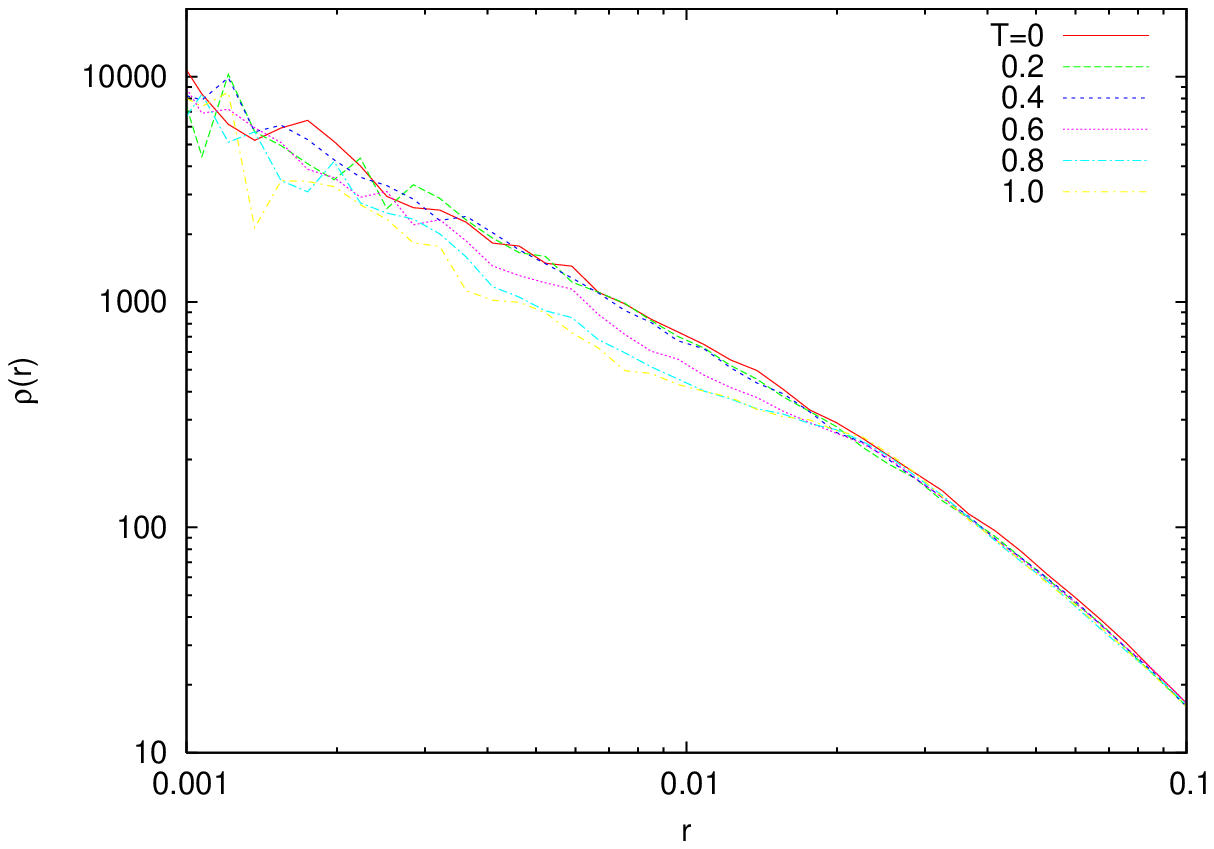}
  \caption{Comparison of evolution in NFW and Hernquist dark haloes
    with matched cusps.  Left: evolution of pattern speed.  Right:
    evolution of profiles in Hernquist halo (compare with Fig.
    \protect{\ref{fig:profiles}}).}
  \label{fig:hern_comp} 
\end{figure}

As a final example, in Figure \ref{fig:hern_comp}, we compare the pattern
speed and density profile evolution of the fiducial bar in a Hernquist
dark halo with that in a NFW halo. However, in this case we match the
densities of the inner profile, making the mass inside of
$R_{vir}$ no longer the same in the two models.  But since the main
differences between these halo models occurs at radii well beyond the bar,
the bar slow down and density profile evolution haloes is nearly identical.

\section{Comparison with other published work}
\label{sec:others}

Recent work from different groups have reported a range of conclusions
regarding halo induced bar slow down and the subsequent bar-induced
halo evolution.  As this and Paper I demonstrate, the theory of the
dynamical interaction agrees with our simulations, demonstrating its
conceptual soundness and applicability.  Any differences must either
be the result of differences in the simulated bar and halo, which can
be in the halo profile, the bar shape, the bar mass, the pattern
speed, or the evolutionary history, or to numerical deficiencies.
\S\ref{sec:ncrit} illustrates the diversity of evolutionary results
that can result from a range of bar and halo models.  We also
demonstrated how not satisfying any of the three particle number
criteria can result in underestimating bar slow down and the halo
density profile evolution.  However, if one uses an insufficient
number of particles in a simulation, how the evolution proceeds
relative to the correct solution will depend on the details of the
N-body algorithm. Furthermore, the approach to the correct answer can
be sudden, which means that the standard approach of convergence
testing can be misleading. Only by fully understanding the underlying
dynamics can one fully trust the results of one's N-body simulation.

There are two main differences reported in the literature: the rate of
bar slowing and the resulting evolution of the halo profile as a
result of the angular momentum deposition.  Most groups agree that the
bar does slow in most circumstances
\citep[HW;][]{Debattista.Sellwood:00, Sellwood:06, Athanassoula:03,
McMillan.Dehnen:05} although \citet[hereafter
VK]{Valenzuela.Klypin:03} find a more modest slow down.  Besides our
work \citep[WK;][]{Holley-Bockelmann.Weinberg.ea:05}, no published
simulation to date has shown a decrease in the halo central density
owing to angular momentum loss from the bar.  Of these,
\citet[hereafter S06]{Sellwood:06} provides sufficiently detailed
information so that we can compare directly.  S06 describes
simulations with an external bar potential similar to those here and
in WK.  His simulations that use an external bar potential are use two
dark matter haloes: a Hernquist model and NFW model.  The bar is a fit
to an $n=2$ Ferrers model \citep{Ferrers:87} also used by HW.  S03
makes two main points relevant to our work here: 1) no profile
evolution is observed for bisymmetric, even-parity forcing by the bar
in either the NFW profile, Hernquist profile or a self-consistently
formed N-body bar; and 2) profile evolution is seen if the odd-parity
forcing is included owing to a centring artifact.  We confirm this
latter artifact, attributable to the dipole response, does distort the
profile, and addressed this issue in detail (see \S\ref{sec:dipole}).
However, our fiducial run (\S\ref{sec:fid}) does not suffer from this
artifact and still shows significant halo density evolution.
Moreover, we find that even in WK, there is profile evolution before
the dipole centring problem dominates the evolution.

There are probably two main reasons for our difference with the
bisymmetric (even terms only) runs from S06.  First, the Ferrers model
fit used in S06 and HW has significantly lower amplitude than the one
used here as described in \S\ref{sec:HWWK}.  Figure \ref{fig:potcof}
shows the fit from HW to Ferrers $n=2$ model compared with our current
model using the same bar mass.  The peak of the quadrupole potential
differs by a factor of 30.  In addition, we have seen that narrower
bars have much larger inner quadrupole potentials than rounder bars
(see Fig.  \ref{fig:quadcomp}).  In particular, S06 and HW use
$b/a=0.5$ while in this paper we use $b/a=0.2$, which is more typical
of a moderate to strong bar.  The inner quadrupole for the same
homogeneous ellipsoid with these two axis ratios is an additional
factor of 10 larger!  When we evolve simulations using the parameters
in S06 we obtain the same results: no measurable halo evolution.  We
expect that similar considerations explain the lack of evolution
reported in S03 from the self-consistent bar run.

More generally, the dynamics described in Paper I and demonstrated here
imply that halo
evolution through bar--halo coupling depends on a variety of factors:
1) the underlying halo and disk mass determines the radial location of
commensurate frequencies; 2) the bar shape and profile determine the
strength of the quadrupole coupling at these resonances; 3) the history of
the evolution, e.g. formation and slow down rate, determines the
duration of the coupling; and 4) all of the previous three determine
the magnitude of the halo evolution.  An interesting example of this
interplay, suggested by the simulations in S03, is the profile
evolution in a Hernquist profile compared to the NFW profile absent
the dipole-centring problems.  A naively analogous fiducial model puts
the bar at the Hernquist model scale length $a$ with the same mass.
Surprisingly, the profile evolution for this model is negligible.
However, the mass enclosed within the bar radius is 2.5 times larger
than that for the NFW profile.  Increasing the bar mass by a factor of two
gives more comparable enclosed masses and the density profile evolves.

VK attribute the differences between their simulations with other work
and analytic estimates to the higher spatial resolution obtainable
with their AMT \citep{Kravtsov.Klypin.ea:97} Poisson solver.  This
conclusion is inconsistent at least in part with the dynamics
described here and in Paper I.  The coupling between the bar and the
halo is dominated by the lowest order even multipoles, $l=2$ in
particular, and the contribution for higher multipoles decreases
exponentially with $l$ beyond some modest value of $l$.  Although VK
use a multimass halo phase space, their effective particle number
falls below the coverage and small-scale noise criteria for the ILR,
although it is just sufficient for the DRR. The failure of one or more
of these criteria may account for some of the discrepancy.
Additionally, it is probably worth investigating the numerical noise
properties of the AMT algorithm. The simulations of
\citet{McMillan.Dehnen:05} also do not have nearly enough particles to
satisfy the coverage criterion for ILR, which is the resonance
responsible for halo cusp evolution.

The underlying dynamics described by \citet{Athanassoula:03} is
consistent with our discussion in Paper I although the emphasis and
numerical experiments are quite different so we will not attempt a
direct comparison here.  However, given the results with the Hernquist
profiles described above and the large scale of her bars relative to
the halo scale length, we would expect that the magnitude of the coupling
to the inner cusp would be small.

\section{Discussion and Summary}
\label{sec:sum}

This paper, together with Paper I, has three goals: 1) to explain and
clarify the dynamics of resonances in secular evolution; 2) to derive
the criteria necessary for a N-body simulation to obtain these resonance
dynamics; and 3) to demonstrate the physics of resonances and the
applicability of these criteria in N-body simulations.  This paper
emphasises the bar--halo interaction as a test case for
resonant-driven secular evolution but the same techniques can be used
generally.  We confirm our prediction \citep[WK]{Weinberg.Katz:02}
that angular momentum transport can drive substantial cusp evolution
within $\approx 30$\% of the bar radius,
mediated by the inner Lindblad resonance.  Our N-body simulations
corroborate the particle-number predictions of Paper I and illustrate
the failure of cusp evolution when the particle criteria are not met.
A small number of low-order resonances dominate the angular momentum
transport.  Therefore, the features and magnitude of the evolution may
vary considerably with the model depending on which resonances exist
and can couple effectively.  More generally, this paper underscores
the importance of a prior understanding of the underlying dynamics to
ensure a successful simulation.

The importance of resonances is easily motivated.  Weak but
large-scale perturbations from satellites, self-excited spiral arms,
and bars break the axisymmetry of a near-equilibrium galaxy.  These
perturbations require resonances to redistribute angular momentum
which drive the secular evolution.  Conversely, without resonances, the
torque would vanish and no evolution could ever be driven by these
perturbations.  For example, the existence of the bar forces an
oscillation in the orbits but many of these orbits remain
adiabatically invariant to the bar forcing.  However, because the bar
is slowing or changing at a finite rate, a measurable number of orbits
end up with broken invariants and cause a net angular momentum change.
However, the change in angular momentum of these orbits varies from
positive to negative with initial phase.  These larger variations
cancel to leave the net positive contribution from the bar. Therefore,
a simulation must have sufficient particles to pick up the net effect
of this cancellation and sufficiently low noise that an orbit remains
coherent over many orbital periods during the resonant interaction.

We have discussed this and several other key points to keep in mind
when considering resonant interactions.  First, Paper I further
elaborates how and why resonances govern the long-term evolution
near-equilibrium of slowly evolving galaxies.  Without resonances,
evolution would depend on astronomical sources of scattering and
scattering has an {\em extremely} long time scale for a quiescent
galaxy.  Second, resonances are broadened in frequency space by both
the finite life time of the bar and the evolution of the bar.
Therefore, no bar resonance is ever infinitely thin in frequency
space, even one with a constant pattern speed.  To first approximation
in the ratio of the evolution time scale to the characteristic period,
the width of the frequency broadening has no effect on the angular
momentum exchange.  We demonstrate this empirically in
\S\ref{sec:barsize} using slowing bars of different sizes and masses.
This result can motivated by an oscillator analogy.  The resonance
acts as a forcing whose frequency is changing with time. The rate of
change for this forcing corresponds to the resonance sweeping through
phase space.  An orbit near resonance has the perturbation imposed on
the natural orbital motion with a beat frequency.  As the orbit
approaches the resonance, this beat frequency decreases and the
perturbation amplitude increases.  The phase of the beat pattern as
the beat frequency passes through zero then determines the torque.  If
the time to pass through the resonance is short, many orbits will be
left with a torque from the bar although the change from each will be
small.  If the time to pass through is long, fewer orbits will receive
a net torque but those that do will receive proportionately more
angular momentum.  In second-order perturbation theory, these two
opposing trends precisely cancel and the torque is the same
independent of waiting time.  In other words, it is the integral under
the frequency ``line'' that matters, not its width.  For example, the
torque transferred in an idealised perturbation with a constant
pattern speed \citep[e.g.][]{Weinberg.Katz:02} will have the same
torque as those with a varying pattern speed over a few dynamical
times.

There are two additional conditions necessary for this to work: the
angular momentum exchange per orbit must remain linear, $|\Delta
J|/J\ll1$, and the rate of change in pattern speed must be larger than
the squared libration frequency for the resonance (the {\em fast
limit} from TW).  This simple explanation is complicated by the
non-linear dynamics that plays a role when the amplitude of the
perturbation is large or the bar evolves slowly.  This is the {\em
slow limit} from TW and the resulting perturbation theory is no longer
second-order.  Paper I shows that this limit is important for the ILR, in
particular, and angular momentum exchange in this regime is very
susceptible to small-scale noise (see Paper I for details).  For
strong bars, many of the low-order resonances are close to the
transition between these two regimes.

These considerations motivated particle-number criteria for N-body
simulations to accurately represent the resonance dynamics.  The
first, the {\em coverage} criterion, ensures that phase space density
be sufficiently high that the sum of interactions with different phase
accurately yields the net torque.  The second and third criteria
ensures that the gravitational potential fluctuations are sufficiently
small that slow-limit resonances can still occur in the simulation.
We artificially divide the noise into two regimes: 1) {\em
small-scale} noise refers to fluctuations on interparticle scales or
larger.  This noise is typical of particle-particle codes; 2) {\em
large-scale} noise refers to fluctuations on the scale of the
inhomogeneity of the galaxy equilibrium.  This noise is typical of the
basis expansion code used in this study.

Different combinations of these three criteria dominate for different
astronomical scenarios and for different codes.  We find here and in
Paper I that the coverage and small-scale noise criteria dominate.
Since our work here uses the basis expansion, our simulations are only
affected by the coverage criterion.  We show that a scale-length bar
(similar to the Milky Way) in an NFW halo with $c=15$ requires $N\gta
2\times10^9$ particles to recover the predicted evolutionary features.
We accomplish this large value of $N$ with a multimass phase-space
distribution (\S\ref{sec:expansion}) which yields a large effective
particle number where it is needed.  We strongly encourage others to
investigate the importance of these criteria and the noise criteria in
particular for their favourite Poisson solver.  It is possible, for
example, that tree and grid codes might introduce sources of noise not
considered here and in Paper I.

The shape and mass of the bar affects these estimates.  A bar with the
same size and shape as our fiducial model but a smaller mass will have a
smaller resonance region and, therefore, require more particles to
satisfy the coverage criterion.  Paper I shows that this criterion
scales approximately as $M_b^{-1/2}$ and our simulations here are in
good agreement with this scaling.  Similarly, a bar with a 5:1
in-plane axis ratio and shallow or flat surface density (similar to
observed strong bars) can provide more than an order of magnitude
stronger coupling that a weak oval with a falling surface density.  We
have illustrated these features with a suite of simulations for bars
with differing axis ratios and different particle numbers.  We find
that the coupling to the inner cusp appears at the bar strength
predicted by our particle number criteria.

Resonance-driven secular evolution is related to dynamical friction,
which is also a secular process.  Traditional Chandrasekhar dynamical
friction is non-periodic and works according to the classical
scattering explanation.  For quasi-periodic systems like galaxies, the
correct physical picture is angular momentum transfer near resonances,
as described in TW and Paper I at length.  Technically, these dynamics
can be understood as the superposition of many second-order secular
Hamiltonian perturbation theory problems, one for each resonance.  In
the limit of a small scale perturbation, say a tiny satellite in orbit
in a dark matter halo, one can show that these two views give
identical results \citep{Weinberg:86}.  So dynamical friction operates
in quasi-periodic systems because of resonances.  This does not imply
that one can replace the dynamics described in Paper I with
Chandrasekhar's dynamical friction formula.  For example, suppose that
we removed the band of phase space around some of the low-order
resonances in Figure 5 in Paper I and filled it in with unresponsive
particles.  The torque would vanish but the torque predicted by
Chandrasekhar's formula would only drop by a negligible amount.  Our
numerical experiment of a rotating bar inside of a King-model core
(\S\ref{sec:profiles}) shows that the torque is negligible even though
the dark-matter density is the same as the rapidly slowing bar in an
NFW model, because too few low-order resonances are available to
accept torque.  A larger bar in the same model may be able to slow
through the outer resonances but, without an ILR, the profile
evolution would be quite modest.  Finally, all halo profiles affect
the magnitude and location of the bar--halo torque through the
existence of various resonances.  For example, by eliminating the ILR
and DRR resonances with a small bar-sized core in the dark-matter
halo, we can all but eliminate subsequent bar slowing.  If, for
example, an early bar could flatten an initially cuspy profile, the
CDM bar slowing \emph{catastrophe} \citep{Debattista.Sellwood:00}
would be eliminated.
   
Paper I derives analytically and demonstrates numerically that noise
from the finite number of particles leads to fluctuations in the
gravitational potential that can change the nature of the resonant
interaction that would obtain in the limit $N\rightarrow\infty$.  In
Nature, galaxies have a variety of substructures and the sum total
influence of these may be described as {\em noise}.  However, all
noise does not have the same spatial and temporal scales as particle
noise and, therefore, will have a different impact on the resonances.  For
example, consider a single satellite orbiting in the halo.  We can
consider its influence on the galaxy using the perturbation theory
described in Paper I and simulations by the methods described in this
paper.  If we add a few more satellites, the net effect can be treated
in the same way as one satellite.  As long as the influence of each
satellite is coherent over particle orbital times, the same resonant
dynamics obtains.  As we begin to increase the number of satellites,
the combined effect will appear as rapid changes in the force in space
and time.  As these fluctuations in the correlation length of the
gravitational field approach the size of the resonance potential and
if the correlation times become smaller than the orbital times, the net
effect on the resonances is the same as with small-scale noise.
However, if we restrict the population of satellites to large radii,
the spatial scales of the fluctuations will be large and the temporal
correlations will be long; therefore, this population only weakly
effects the dynamics of the inner galaxy.  As another example,
molecular clouds in a disk, because they are confined to the disk,
only contribute noise on scales that are small relative to the
dominant halo resonances considered in Paper I.  Noise owing to the finite
number of simulation particles generates noise at all scales, and
because this noise source is generated by the orbits themselves, it
naturally couples to the spatial and temporal noise scales in the most
effective combination to affect the resonances.  In summary, all noise
is {\em not} alike; the spatial and temporal properties of the noise are
just as important as its total power.  The most obvious sources of
astronomical noise---molecular clouds in the disk and substructure in the
halo---are not likely to have the same effect as particle noise.  The
power on various scales for some particular noise source may be
computed using the method described in \citet{Weinberg:01b}.

Finally these stringent particle number requirements does not bode
well for {\it ab initio} galaxy formation simulations
(e.g. Steinmetz.Navarro:XX and Governato.etal:05) to correctly capture
the correct resonant dynamics.  Without doing so, results about disk
sizes and rotation curves, some of the most controversial topics, are
all suspect. For example, a typical such simulation that focuses on
the formation of one galaxy requires about five times more particles
than those required within the virial radius to place the galaxy in
its proper cosmological context.  Such simulations also use equal mass
particles within the virial radius.  This means that such a simulation
would require at least 10 billion particles to correctly model the
disk dynamics! This is just to model the bar-halo interaction. Other
processes would likely require even more particles.  Such large
simulations seem unlikely with our lifetime and hence astronomers will
have to invent more clever approaches than pure brute force to make
progress in this area of study.

This study motivates the importance of combining simulations with
analytic theory to understand galaxy evolution.  To recapitulate the
introduction, the bar--halo problem was chosen for its overall
structural simplicity and, therefore, its amenability to both
perturbation theory and N-body simulation.  To our surprise, the
problem exhibited subtlety at every turn.  From perturbation theory,
we find that two separate regimes apply: one that scales as the square
of the perturbation strength and one that scales as the square root of
the perturbation strength.  Both regimes apply to problems of
astronomical interest.  From kinetic theory, we find that these
regimes can be affected by relaxation for particle numbers currently
used by simulations. We further show that the quantity of torque and
the qualitative behaviour of the subsequent evolution depends on most
aspects of the model including its history.  Because of these
subtleties, N-body studies have produced varying results leading to
controversy in the literature.  Moreover, researchers typically rely
on N-body simulation to study astronomical scenarios that appear
trickier than this one.  For example, the heating of a disk by a satellite
may require proper representation of bending modes and subsequent
damping through resonant coupling.  The range of frequencies and
scales in this latter problem is much broader than the bar--halo
interaction considered here and, therefore, likely to present similar
sorts of surprises that only a combined N-body--perturbation theory
comparison will reveal.  Additional multiscale problems include the
satellite--halo interaction and resonant (pseudo-)bulge formation, and
these are likely to require similarly intense levels of dynamical
scrutiny.

\section*{Acknowledgments}

Many thanks to Kelly Holley-Bockelmann for suggestions, discussions
and a careful reading of this manuscript.  MDW would also like to
acknowledge many electronic discussions with Jerry Sellwood.  MDW
completed the early stages of this work at the Institute for Advanced
Study in Princeton and thanks his host, John Bahcall, for his
hospitality.  This work was supported in part by NSF AST-0205969 and
AST-9988146 and by NASA ATP NAGS-13308 and NAG5-12038.

\label{lastpage}

\end{document}